\documentclass[twocolumn,dvipsnames]{aastex63}
\usepackage{gensymb}
\usepackage{amsmath}
\usepackage{amssymb}
\usepackage{xcolor}
\usepackage{physics}
\usepackage{bm}
\usepackage{bigints}
\usepackage{subfigure}

\usepackage{pifont}

\hypersetup{colorlinks=true, linkcolor=red, filecolor=magenta, urlcolor=Sepia, citecolor=blue}

\definecolor{ForestGreen}{HTML}{228B22}

\received{XXX}
\revised{YYY}
\accepted{ZZZ}
\submitjournal{ApJ}

\shorttitle{Gravitational wave in $f(R)$ gravity}
\shortauthors{Kalita \& Mukhopadhyay}

\begin{document}

\title{Gravitational wave in $f(R)$ gravity: possible signature of sub- and super-Chandrasekhar limiting mass white dwarfs}


\author[0000-0002-3818-6037]{Surajit Kalita}
\affiliation{Department of Physics, Indian Institute of Science, Bangalore 560012, India}
\email{surajitk@iisc.ac.in}

\author[0000-0002-3020-9513]{Banibrata Mukhopadhyay}
\email{bm@iisc.ac.in}
\affiliation{Department of Physics, Indian Institute of Science, Bangalore 560012, India}

\begin{abstract}
After the prediction of many sub- and super-Chandrasekhar (at least a dozen for the latter) limiting mass white dwarfs, hence apparently peculiar class of white dwarfs, from the observations of luminosity of type Ia supernovae, researchers have proposed various models to explain these two classes of white dwarfs separately. We earlier showed that these two peculiar classes of white dwarfs, along with the regular white dwarfs, can be explained by a single form of the $f(R)$ gravity, whose effect is significant only in the high-density regime, and it almost vanishes in the low-density regime. However, since there is no direct detection of such white dwarfs, it is difficult to single out one specific theory from the zoo of modified theories of gravity. We discuss the possibility of direct detection of such white dwarfs in gravitational wave astronomy. It is well-known that in $f(R)$ gravity, more than two polarization modes are present. We estimate the amplitudes of all the relevant modes for the peculiar as well as the regular white dwarfs. We further discuss the possibility of their detections through future-based gravitational wave detectors, such as LISA, ALIA, DECIGO, BBO, or Einstein Telescope, and thereby put constraints or rule out various modified theories of gravity. This exploration links the theory with possible observations through gravitational wave in $f(R)$ gravity.
\end{abstract}


\section{Introduction}\label{Introduction}

White dwarfs (WDs) are the end-state of stars with mass $\lesssim 8 M_\odot$. A WD attains its stable equilibrium configuration by 
balancing the outward force due to the degenerate electron gas with the inward force of gravity. If the WD has a binary companion, it 
pulls out matter from the companion, and as a result, the mass of WD increases. Once the mass of the WD reaches the Chandrasekhar 
mass-limit \citep{1931ApJ....74...81C} ($\sim 1.4 M_\odot$ for carbon-oxygen non-rotating non-magnetized WDs), the pressure balance no 
longer sustains, and the WD bursts out to produce a type Ia supernova (SNIa) \citep{choudhuri_2010}. The similarity in peak luminosities 
of SNeIa is used as one of the standard candles to estimate the luminosity distances for various astronomical and cosmological objects 
\citep{1987ApJ...323..140L,1997Sci...276.1378N}. However, recent discoveries of various under- and over-luminous SNeIa question the 
complete validity of considering luminosities of SNeIa as standard candle. SNeIa such as SN 1991bg \citep{1992AJ....104.1543F,1997MNRAS.284..151M}, SN 1997cn \citep{1998AJ....116.2431T}, SN 1998de \citep{2001PASP..113..308M}, SN 1999by \citep{2004ApJ...613.1120G}, and SN 2005bl \citep{2008MNRAS.385...75T} 
were discovered with extremely low luminosities, which were produced from WDs with $^{56}$Ni mass content as low as $\sim 
0.1 M_\odot$ \citep{2006A&A...460..793S}. On the other hand, a different class of SNeIa, such as, SN 2003fg \citep{2006Natur.443..308H}, SN 2006gz \citep{2007ApJ...669L..17H}, SN 2009dc \citep{2009ApJ...707L.118Y,2010ApJ...714.1209T,2011MNRAS.410..585S,2011MNRAS.412.2735T,2012ApJ...756..191K}, 
SN 2007if \citep{2010ApJ...713.1073S,2010ApJ...715.1338Y,2012ApJ...757...12S}, SN 2013cv \citep{2016ApJ...823..147C}, any many more 
was discovered with an excess luminosity, with the observed mass of $^{56}$Ni as high as $\sim 1.8 M_\odot$ \citep{2012ApJ...756..191K}, 
violating the Khokhlov pure detonation limit \citep{1993A&A...270..223K}. It was inferred that these under-luminous SNeIa were produced 
from WDs with mass $\sim 0.6 M_\odot$ \citep{1997MNRAS.284..151M,1998AJ....116.2431T}, while the same for over-luminous SNeIa could be 
$\sim 2.8 M_\odot$ \citep{2010ApJ...713.1073S,2012ApJ...756..191K}. Hence. these progenitor WDs of peculiar SNeIa violate the 
Chandrasekhar mass-limit: the under-luminous SNeIa were produced from sub-Chandrasekhar limiting mass WDs (WDs burst before reaching the 
mass $\sim 1.4 M_\odot$), and the over-luminous SNeIa were produced from super-Chandrasekhar limiting mass WDs (WDs burst well above the 
mass $\sim 1.4 M_\odot$). These new mass-limits are important as they may lead to modifying the standard candle.

Various groups around the world have proposed different models to explain the formation of these two peculiar classes of SNeIa. 
Sub-Chandrasekhar limiting mass WDs were believed to be formed by merging two sub-Chandrasekhar mass WDs (double degenerate scenario) 
leading to another sub-Chandrasekhar mass WD, exploding due to accretion of a helium layer \citep{2010Natur.463...61P,2000ARA&A..38..191H}. 
On the other hand, the super-Chandrasekhar WDs were often explained by incorporating different physics, such as a double degenerate 
scenario \citep{2007ApJ...669L..17H}, presence of magnetic fields \citep{2013PhRvL.110g1102D,2014JCAP...06..050D}, presence of a 
differential rotation \citep{2012ApJ...744...69H}, presence of charge in the WDs \citep{2014PhRvD..89j4043L}, 
ungravity effect \citep{2016PhRvD..93j4046B}, lepton number violation in magnetized white dwarfs \citep{2015NuPhA.937...17B}, generalized Heisenberg uncertainty principle \citep{2018JCAP...09..015O}, and many more. However, 
none of these theories can self-consistently explain both the peculiar classes of WDs. Moreover, each of these has some caveats or
incompleteness, mostly based on the stability \citep{1989MNRAS.237..355K, 2009MNRAS.397..763B}. Furthermore, numerical simulations 
showed that a merger of two massive WDs could never lead to the mass as high as $2.8M_\odot$ due to the off-center ignition, and 
formation of a neutron star rather than an (over-luminous) SNIa \citep{2004ApJ...615..444S,2006MNRAS.373..263M}. Hence, all the 
conventional pictures failed to explain the inferred masses of both 
the sub- and super-Chandrasekhar progenitor WDs and also both the 
classes of progenitor WDs simultaneously by invoking the same 
physics. Moreover, each of the theories can explain only one regime of SNIa but it seems more likely that the nature would prefer 
only one scenario/physics to exhibit the same class of supernovae. 
Whether it be an under- or over-luminous SNeIa, other physics such 
as the presence of Si etc. remains the same. Therefore, we seem to 
require just one theory to explain all the SNeIa.

Einstein's theory of general relativity (GR) is undoubtedly the most beautiful theory to explain the theory of gravity. It can easily 
explain a large number of phenomena where the Newtonian gravity falls short, such as the deflection of light in strong gravity, 
generation of the gravitational wave (GW) in $3+1$ dimension, perihelion precession of Mercury's orbit, gravitational redshift of 
light, to mention a few. It is well-known that in the asymptotically flat limit where the typical velocity is much small compared to the 
speed of light, GR reduces to the Newtonian theory of gravity \citep{2009igr..book.....R}. According to the Newtonian theory, 
Chandrasekhar mass-limit for WDs is achieved only at zero radius with infinite density, whereas GR can consistently explain this for a 
WD with a finite radius and a finite density \citep{padmanabhan_2001}. Nevertheless, following several recent observations in cosmology 
\citep{2016ARNPS..66...95J,2017PDU....18...73C}
and in the high-density regions of the universe, such as at the vicinity of compact objects 
\citep{2019JCAP...06..029H,2020PhRvD.101d1301B,2020PhRvD.101j4057B,2020arXiv200804404M},
it seems that GR may not be the ultimate 
theory of gravity. \cite{1980PhLB...91...99S} first used one of the modified theories of gravity, namely $R^2$-gravity with 
$R$ being the scalar curvature, to explain the cosmology of the very early universe. Eventually, researchers have proposed a large 
number of modifications to GR, e.g., various $f(R)$ gravity models, to elaborate the physics of the different astronomical systems, such 
as the massive neutron stars \citep{2013JCAP...12..040A,2014PhRvD..89j3509A}, accretion disk around the compact object 
\citep{2006PhRvD..74f4022M,2008PhRvD..78b4043P,2013A&A...551A...4P,2019EPJC...79..877K} and many more.
Our group has also shown that 
using the suitable forms of $f(R)$ gravity, we can unify the physics of all WDs, including those possessing sub- and super-Chandrasekhar 
limiting masses \citep{2015JCAP...05..045D,2018JCAP...09..007K}. We showed that by fixing the parameters in a viable $f(R)$ gravity 
model such that it satisfies the solar system test \citep{2014IJMPD..2350036G}, one can obtain the sub-Chandrasekhar limiting mass WDs 
at a relatively low density and super-Chandrasekhar WDs at high density \citep{2018JCAP...09..007K}. Of course, the mass--radius 
relation alters from Chandrasekhar's original mass--radius relation depending on the form of $f(R)$ gravity model. 
Nevertheless, from the recent detection of GW through LIGO/Virgo detectors, researchers have put 
constraints on the $f(R)$ gravity theory \citep{2019PhRvD..99d4056J,2017GReGr..49...99V}.

It is important to note that all the inferences of sub- and super-Chandrasekhar limiting mass WDs were made indirectly from the 
luminosity observations of SNeIa. There is, so far, no direct detection of super-Chandrasekhar WDs in any electromagnetic surveys such 
as GAIA, Kepler, SDSS, or WISE, as the massive WDs are usually less luminous compared to the lighter ones \citep{2018MNRAS.477.2705B,2020MNRAS.496..894G}. 
Therefore, the exact sizes of these peculiar WDs are unknown and, hence, nobody could, so far, single out the exact theory of gravity 
from the various probable models. On the other hand, recent detection of GWs from the merger events opens a new window in astronomy. 
Since a strong magnetic field and high rotation can increase the mass of a WD, such a WD, possessing a 
specific configuration, can emit GW efficiently \citep{2019MNRAS.490.2692K,2020IAUS..357...79K,2020MNRAS.492.5949S}. Various advanced futuristic GW 
detectors such as LISA, DECIGO, and BBO can detect this gravitational radiation for a long time, depending on the field geometry and its 
strength \citep{2020ApJ...896...69K} and, thereby, one can estimate the size of the super-Chandrasekhar WDs, where the electromagnetic 
surveys were not successful.

Since $f(R)$ gravity is a better bet to explain and unify all the WDs, including the peculiar ones, to study its validity from 
observation is extremely necessary. Due to the failure of their direct detections in electromagnetic surveys, GW astronomy seems to be 
the prominent alternate to detect the peculiar WDs directly. In this way, one can estimate both the mass and the size of the objects, 
thereby ruling out or putting constraints on the various models of $f(R)$ gravity. Moreover, there is a huge debate on the existence of 
modifications to GR. Hence, if the futuristic GW detectors such as LISA, ALIA, DECIGO, BBO, or Einstein Telescope can detect 
such WDs, it will also be a simple verification for the existence of the modified theories of gravity. We, in this article, present 
various mechanisms that can produce GWs from $f(R)$ gravity induced WDs, and discuss how to rule out/single out various theories from 
such observations. 

The article is organized as follows. In \S \ref{GW in f(R) gravity}, we discuss the properties of gravitational radiation 
in $f(R)$ gravity. In \S \ref{Mechanisms for generating GW}, we discuss the generation of GW from $f(R)$ gravity induced WDs through 
various mechanisms such as the presence of roughness at the surface of WDs, or the presence of magnetic fields and rotation in the 
WDs. In \S \ref{Strength of GW emitted from an isolated WD}, we discuss the amplitude and luminosity of the gravitational radiation 
emitted from these isolated WDs, and whether the futuristic detector can detect them or not. In \S \ref{Discussion}, we discuss the 
various results and their physical interpretations. In this section, we mainly discuss how to extract information about the WDs from GW 
detection, and thereby to put constraints on the modified theories of gravity, before we conclude in \S \ref{Conclusions}.


\section{Gravitational wave in $\lowercase{f}(R)$ gravity}\label{GW in f(R) gravity}

Assuming the metric signature to be $(-,+,+,+)$ in four dimensions, the action in $f(R)$ gravity (modified Einstein-Hilbert action) is 
given by \citep{2010LRR....13....3D,2014LRR....17....4W,2017PhR...692....1N}
\begin{align}
\mathcal{S}_{f(R)}=\int\left[\frac{c^3}{16 \pi G}f(R)+\mathcal{L}_\mathcal{M}\right]\sqrt{-g}\dd[4]{x},
\end{align}
where $c$ is the speed of light, $G$ Newton's gravitational constant, $\mathcal{L}_\mathcal{M}$ the Lagrangian of the matter field and 
$g=\det(g_{\mu\nu})$ the determinant of the spacetime metric $g_{\mu\nu}$. Varying this action with respect to $g_{\mu\nu}$, with 
appropriate boundary conditions, we obtain the modified Einstein equation in $f(R)$ gravity, given by
\begin{equation}\label{modified equation}
F(R)R_{\mu \nu}-\frac{1}{2}g_{\mu \nu}f(R) - \left(\nabla_\mu \nabla_\nu-g_{\mu \nu}\Box \right)F(R) = \kappa T_{\mu \nu},
\end{equation}
where $R_{\mu\nu}$ is the Ricci tensor, $T_{\mu\nu}$ the matter stress-energy tensor, $F(R) = \dv*{f(R)}{R}$, $\kappa = 8\pi G/c^4$, $\Box$ 
the d'Alembertian operator given by $\Box = -\partial_t^2/c^2 + \laplacian$ with $\partial_t$ being the temporal partial derivative and 
$\laplacian$ the 3-dimensional Laplacian. For $f(R)=R$, it is obvious that Equation \eqref{modified equation} will reduce to the 
field equation in GR \citep{2009igr..book.....R}. The trace of Equation \eqref{modified equation} is given by
\begin{equation}\label{modified trace equation}
RF(R)-2f(R)+3\Box F(R) = \kappa g^{\mu \nu} T_{\mu \nu}  = \kappa T.
\end{equation}
Since we plan to explore $f(R)$ gravity models, which can explain both the sub- and super-Chandrasekhar limiting mass WDs together, the 
first higher-order correction to GR, i.e., $f(R) = R + \alpha R^2$ seems to suffice for this purpose. However, in this model, one needs 
to vary the model parameter $\alpha$ to obtain both the regimes of the WDs \citep{2015JCAP...05..045D} and, hence, this is probably not 
the best model for this purpose. Therefore, we need to consider the next higher order correction terms, i.e., 
$f(R) = R + \alpha R^2 \{1 - \gamma R  + \mathcal{O}(R^2)\}$ \citep{2018JCAP...09..007K}, to remove the deficiency of the previous model. In this 
model, one does not need to vary the parameters $\alpha$ and $\gamma$ present in the model. Rather one needs to fix them from the 
Gravity Probe B experiment and, then, just changing the central density, one can obtain both the sub- and super-Chandrasekhar limiting 
mass WDs. \cite{2018JCAP...09..007K} provided a detailed analysis of this considering various higher order 
corrections to GR and establishing that they still pass the solar system test.

Since we are interested in the most common scenario where GW propagates in vacuum, i.e. in the flat Minkowski space-time, we need to 
linearize both $g_{\mu\nu}$ as well as $R$. The perturbed forms for $g_{\mu\nu}$ and $R$ are given by
\begin{align}
g_{\mu\nu} &= \eta_{\mu\nu} + h_{\mu\nu}, \\
R &= R_0 + R_1,
\end{align}
with $\abs{h_{\mu\nu}} \ll \abs{\eta_{\mu\nu}}$, where $\eta_{\mu\nu}$ is the background Minkowski metric, $R_0$ the unperturbed background 
scalar curvature, 
and $h_{\mu\nu}$ and $R_1$ are respectively the tensor and scalar perturbations. Of course, for Minkowski vacuum background, $R_0 = 0$ 
and $T_0=0$, where $T_0$ being the trace of the background stress-energy tensor. Now perturbing the equations \eqref{modified equation} 
and \eqref{modified trace equation}, and substituting the above relations, we obtain the linearized field equations, given by 
\citep{2008PhLB..669..255C,2017PhRvD..95j4034L}
\begin{align} \label{Eq: perturbed equations: tensor}
\Box \bar{h}_{\mu\nu} &= -\frac{16 \pi G}{c^4}T_{\mu\nu} \\ \label{Eq: perturbed equations: scalar}
\Box h_f - m^2 h_f &= \frac{8 \pi G}{3 F(R_0) c^4} T,
\end{align}
where $\bar{h}_{\mu\nu} = h_{\mu\nu} - \left(h/2 - h_f \right)\eta_{\mu\nu}$ with $h=\eta_{\mu\nu}h^{\mu\nu}$ and $h_f = F'(R_0)R_1/F(R_0)$. 
Here, $m$ is the effective mass associated with the scalar degree of freedom in $f(R)$ gravity \citep{2010RvMP...82..451S,2014IJMPD..2350037P,2019PhRvD..99j4046S}, given by 
\begin{equation} \label{Eq: effective mass}
m^2 = \frac{1}{3}\left\{\frac{F(R_0)}{F'(R_0)}-R_0 \right\},
\end{equation}
where $F' = \dv*{F}{R}$. Of course, in Minkowski vacuum background, $m^2 = 1/3\{F(0)/F'(0)\}$. It is evident that $m$ depends on the 
background density, which is known as the chameleon mechanism in $f(R)$ gravity \citep{2018PhLB..777..286L,2018LRR....21....1B}. In 
Minkowski vacuum background, for $f(R) = R + \alpha R^2 (1 - \gamma R)$ with $F(R) = 1 + \alpha R (2-3\gamma R)$, the equations 
\eqref{Eq: perturbed equations: tensor} and \eqref{Eq: perturbed equations: scalar} reduce to
\begin{align} \label{Eq: tensor mode}
\Box \bar{h}_{\mu\nu} &= -\frac{16 \pi G}{c^4}T_{\mu\nu} \\ \label{Eq: scalar mode}
\left(\Box - m^2\right) R_1 &= \frac{4 \pi G}{3\alpha c^4} T,
\end{align}
with $m^2 = 1/6\alpha$. When the GW propagates in the vacuum, these equations reduce to \citep{2017PhRvD..95j4034L,2008PhLB..669..255C}
\begin{align} \label{Eq: vacuum equations}
\Box \bar{h}_{\mu\nu}  = 0, \qquad
\left(\Box - m^2\right) R_1 = 0.
\end{align}
The first equation is similar to the equation obtained in GR, which means $\bar{h}_{\mu\nu}$ satisfies the transverse-traceless (TT) 
gauge condition. This leads to the fact that there is a presence of only two propagating degrees of freedom/polarization 
for $\bar{h}_{\mu\nu}$ (namely $\bar{h}_+$ and $\bar{h}_\times$). In GR, since $\alpha\to 0$, or equivalently $m\to \infty$, only the 
tensor equation gives the propagating polarization modes. The scalar mode, being infinitely massive in GR, can no longer propagate and, 
hence, the corresponding scalar equation serves as a constraint equation for the tensor modes\footnote{We can also investigate the other extreme regime, i.e., $m\rightarrow 0$ or $\alpha \to \infty$. From Gravity Probe B experiment, the bound on $\alpha$ is $\abs{\alpha}\lesssim 5 \times 10^{15}$ cm$^2$ \citep{2010PhRvD..81j4003N}. It is evident that $\alpha\to\infty$ naturally violates this bound, and hence, this limit is unphysical in the present context.}.
On the other hand, in $f(R)$ gravity, since $R \neq 0$ (even in vacuum, due to the presence of $R_1$), there is a presence of an extra 
propagating scalar degree of polarization, also known as the breathing mode. Hence the number of polarizations in $f(R)$ gravity turns 
out to be 3, unlike the case for GR where it is 2 \citep{2016PhRvD..93l4071K,2018Univ....4...85G}. For a plane wave traveling in 
$z-$direction, the solutions of the above wave Equations \eqref{Eq: vacuum equations}, are given by
\begin{align}
\bar{h}_{\mu\nu}(z,t) &= \hat{h}_{\mu\nu} \exp\left[i (\omega z/c - \omega t) \right], \\
R_1(z,t) &= \hat{R}_1 \exp\left[i \left(\sqrt{\tilde{\omega}^2 - m^2c^2} z/c - \tilde{\omega} t \right) \right],
\end{align}
where $\omega$ is the frequency of the tensorial modes and $\tilde{\omega}$ is that of the scalar mode. It is evident that the tensorial 
modes, being massless, propagate at a speed $v_t=c$, whereas the massive scalar mode propagates with a group velocity 
$v_s = c\sqrt{\tilde{\omega}^2 - m^2c^2}/\tilde{\omega}<c$ \citep{2011JCAP...08..029Y}.

Our aim is to calculate the strength of GW generated from $f(R)$ gravity induced WDs. Hence, we need to solve Equations 
\eqref{Eq: tensor mode} and \eqref{Eq: scalar mode}. Since these are inhomogeneous differential equations, we use the method of Green's 
function. The Green's function for the operator $(\Box - m^2)$ or $(-\partial_t^2/c^2 + \laplacian -m^2)$ is given by \citep{2011PhRvD..83j4022B,2019GReGr..51...84D}
\begin{equation}
\mathcal{G}_m(x,x') = \int \frac{\dd[4]{p}}{(2\pi)^4} \frac{\exp\left[ip.(x-x')\right]}{m^2+p^2},
\end{equation}
where $p \equiv \left(\omega/c, \bm{k}\right)$ with $\bm{k}$ being the wavenumber, such that $p^2=-\omega^2/c^2+\bm{k}^2$. For a 
spherically symmetric system with $x \equiv \left(ct,r=|\bm{x}-\bm{x}'|,\theta,\phi \right)$, where $\bm{x}'$ and $\bm{x}$ are 
respectively source and observer (detector) positions, it reduces to
\begin{equation}\label{Eq: Spherical Green function}
\mathcal{G}_m(x,x') =
\begin{cases}
\bigintsss \frac{\dd{\omega}}{2\pi c} \exp\left[-i\omega(t-t')\right] \frac{1}{4\pi r} \exp\left[i\sqrt{\frac{\omega^2}{c^2} - m^2}~r\right] \\ \hspace{5cm}\text{if $\omega^2 > m^2c^2$}\\
\bigintsss \frac{\dd{\omega}}{2\pi c} \exp\left[-i\omega(t-t')\right] \frac{1}{4\pi r} \exp\left[-\sqrt{m^2 - \frac{\omega^2}{c^2}}~r\right] \\ \hspace{5cm}\text{if $\omega^2 < m^2c^2$}.
\end{cases}
\end{equation}
Similarly, the Green's function for a spherically symmetric $\Box$ operator is given by \citep{2011PhRvD..83j4022B}
\begin{equation}
\mathcal{G}_0(x,x') = \frac{\var (t-t'-r/c)}{4\pi r c}.
\end{equation}
Therefore, from Equation \eqref{Eq: tensor mode}, the solution for $\bar{h}_{\mu\nu}$ is given by
\begin{equation}
\bar{h}_{\mu\nu} = -\frac{4 G}{c^4} \int \dd[3]{x'} \frac{T_{\mu\nu}(t-r,\bm{x'})}{r}.
\end{equation}
Since $\bar{h}_{\mu\nu}$ follows TT gauge condition like GR, only the space part of the $\bar{h}_{\mu\nu}$ contributes. Assuming the 
detector to be far from the source such that $\bm{x}\gg\bm{x}'$, the above equation reduces to the following form \citep{2009igr..book.....R}
\begin{equation} \label{Eq: Linearized tensorial amplitude}
\bar{h}_{ij} = -\frac{2 G}{c^4 r} \ddot{Q}_{ij},
\end{equation}
where $Q_{ij}$ is the quadrupolar moment of the system with $i,j=1,2,3$. Moreover, from Equation \eqref{Eq: scalar mode}, the solution for $R_1$ is given by
\begin{equation} \label{Eq: Linearized Ricci scalar}
R_1 = \frac{4\pi G}{3\alpha c^4} \int \dd[4]{x'} \mathcal{G}_m(x,x') T(x').
\end{equation}
The stress-energy tensor for a perfect fluid is given by
\begin{align}
T^{\mu\nu} = (\rho c^2 + \mathcal{P})u^\mu u^\nu + \mathcal{P} g^{\mu\nu},
\end{align}
where $\mathcal{P}$ is the pressure, $\rho$ the density and $u^\mu$ the four-velocity of the fluid. The trace of $T^{\mu\nu}$ is 
given by $T = -\rho c^2 + 3\mathcal{P}$. For the case of WD, the equation of state (EoS), known as Chandrasekhar EoS, is governed by 
the degenerate electron gas. It is given by \citep{1935MNRAS..95..207C}
\begin{equation}\label{Eq: Chandrasekhar EoS}
\begin{aligned}
\mathcal{P} &= \frac{m_e^4 c^5}{24\pi^2\hbar^3} \left[x_F(2x_F^2-3)\sqrt{x_F^2+1}+3\sinh^{-1}x_F \right],\\
\rho &= \frac{\mu_e m_H(m_ec)^3}{3\pi^2\hbar^3}x_F^3,
\end{aligned}
\end{equation}
where $x_F = p_F/m_ec$, $p_F$ is the Fermi momentum, $m_e$ the mass of an electron, $\hbar$ the reduced Planck's constant, $\mu_e$ the mean 
molecular weight per electron and $m_H$ the mass of a hydrogen atom. For our work, we choose $\mu_e=2$ indicating the carbon-oxygen WDs. 
It is evident, from this EoS, that $\rho c^2 \gg \mathcal{P}$ and, hence, $T \approx -\rho c^2$. Moreover, from the Gravity Probe B 
experiment,the bound on $\alpha$ is $|\alpha| \lesssim 5\times10^{15}$ cm$^2$ \citep{2010PhRvD..81j4003N}. We already showed in our 
previous work that $\alpha = 3\times 10^{14}$ cm$^2$ is enough to probe both sub- and super-Chandrasekhar limiting mass WDs 
simultaneously \citep{2018JCAP...09..007K}. For this value of $\alpha$, in vacuum background, $m^2 = 1/6\alpha \approx 5.6 \times 10^{-16}$ cm$^{-2}$ and, hence, the cut-off frequency turns out to be $\omega_c = mc \approx 709.5$ rad s$^{-1}$. The rotation period 
$\Omega_\text{rot}\sim\omega$ of a WD is always $\lesssim 10$ rad s$^{-1}$. Hence, in Equation \eqref{Eq: Spherical Green function}, 
$\omega^2\ll m^2c^2=\omega_c^2$ is satisfied, and the Green's function reduces to \citep{2010PhRvD..82f4021S}
\begin{equation}
\mathcal{G}_m(x,x') = \frac{\var(t-t')}{4\pi rc}e^{-mr}.
\end{equation}
Therefore, using Equation \eqref{Eq: Linearized Ricci scalar}, for a WD, the solution for $R_1$ is given by
\begin{equation}
R_1 = \frac{G}{3\alpha c^2} \int \dd[3]{x'} \frac{\rho(x')}{\abs{x-x'}}e^{-m\abs{x-x'}}.
\end{equation}
The typical distance of a WD from the earth is $\sim 100$ pc (1 pc $\approx 3.1 \times 10^{18}$ cm), which means $x\gg x'$. Hence, for a 
WD with mass $M$, $R_1$ reduces to
\begin{equation}
R_1 \propto \frac{G M}{3\alpha c^2 r} e^{-mr}.
\end{equation}
For the chosen values of $m$ and $r$, $mr \approx 7.3 \times 10^{12}$, which means the scalar mode's amplitude is exponentially 
suppressed enormously \citep{2019PhRvD..99l4050K}, and the detectors cannot detect them. Hence, in the rest of the article, 
we discuss only the tensorial modes $\bar{h}_{ij}$ given by Equation \eqref{Eq: Linearized tensorial amplitude}, and for convenience, we 
remove `bar' from $h$ hereinafter.


\section{Mechanisms for generating GW from an isolated WD} \label{Mechanisms for generating GW}

In this section, we discuss a couple of mechanisms that can generate gravitational radiation from an isolated WD. From Equation 
\eqref{Eq: Linearized tensorial amplitude}, it is evident that a system can emit gravitational radiation if and only if the system 
possesses a time-varying quadrupolar moment such that $\ddot{Q}_{ij} \neq 0$. Hence, neither a spherically symmetric system nor an 
axially symmetric system can radiate GW, and a tri-axial system is required. In a tri-axial system, the moment of inertia is different 
along all the three spatially perpendicular axes. There are mainly two ways by which a WD can be tri-axial \citep{compact}. First, the 
rotating WD already possesses roughness at its surface, may be due to the presence of mountains and holes (craters). The second 
possibility being that the WD possesses the magnetic field and rotation, with a non-zero angle between their respective axes. We now 
discuss each of these possibilities one by one.


\subsection{GW due to roughness of the surface}

If a WD possesses asymmetry of matter at its surface, the moments of inertia of the WD are different along all directions, making the 
system a tri-axial one. If such a WD rotates with angular velocity $\Omega_\text{rot}$, it can emit effective gravitational radiation 
continuously. Suppose the moments of inertia for such a system be $I_1$, $I_2$ and $I_3$ along $x-$, $y-$, $z-$axes respectively such 
that $I_1 < I_2 < I_3$. Thereby, using Equation \eqref{Eq: Linearized tensorial amplitude}, the two tensorial polarizations of GW, 
at time $t$, are given by \citep{1980PhRvD..21..891Z,2005CQGra..22.1825V,Maggiore}
\begin{equation}\label{Eq: GW polarization}
\begin{aligned}
h_+ &= A_{+,0}\cos\left(2\Omega_\text{rot} t\right) + A_{+,1}\cos\left[{(\Omega_\text{rot}+\Omega_p) t}\right] \\ &+ A_{+,2}\cos\left[{2(\Omega_\text{rot}+\Omega_p) t}\right], \\
h_\times &= A_{\times,0}\sin\left(2\Omega_\text{rot} t\right) + A_{\times,1}\sin\left[{(\Omega_\text{rot}+\Omega_p) t}\right] \\ &+ A_{\times,2}\sin\left[{2(\Omega_\text{rot}+\Omega_p) t}\right],
\end{aligned}
\end{equation}
where
\begin{equation}\label{Eq: GW polarization amplitude}
\begin{aligned}
A_{+,0} &= \left(h_0/2\right)\left(1+\cos^2 i\right),\\
A_{+,1} &= 2h_0'\left(I_1a/I_3b\right)\sin i \cos i,\\
A_{+,2} &= 2h_0'\left(I_1a/I_3b\right)^2\left(1+\cos^2 i\right),\\
A_{\times,0} &= h_0\cos i,\\
A_{\times,1} &= 2h_0'\left(I_1a/I_3b\right)\sin i,\\
A_{\times,2} &= 4h_0'\left(I_1a/I_3b\right)^2\cos i,
\end{aligned}
\end{equation}
with $i$ being the inclination angle between the rotation axis of the WD and the detector's line of sight, and
\begin{equation}\label{Eq: h0 mountain}
\begin{aligned}
h_0 &= -\frac{4G\Omega_\text{rot}^2 I_3}{rc^4} \left(\frac{I_1-I_2}{I_3}\right),\\
h_0' &= -\frac{G\left(\Omega_\text{rot}+\Omega_p\right)^2 I_3}{rc^4} \left(1-\frac{I_1+I_2}{2I_3}\right).
\end{aligned}
\end{equation}
Here $a$ and $b$ are given by
\begin{align}
a = \sqrt{\frac{2\tilde{E}I_3-\tilde{M}^2}{I_1\left(I_3-I_1\right)}}, \qquad
b = \sqrt{\frac{\tilde{M}^2-2\tilde{E}I_1}{I_3\left(I_3-I_1\right)}},
\end{align}
where $\tilde{E} = \left(I_1 \Omega_1^2 + I_2 \Omega_2^2 + I_3 \Omega_3^2\right)/2$ and $\tilde{M}^2 = I_1^2 \Omega_1^2 + I_2^2 \Omega_2^2 + I_3^2 \Omega_3^2$, 
with $\Omega_1$, $\Omega_2$, $\Omega_3$ are the components of initial angular velocity along $x-$, $y-$, $z-$axes respectively \citep{LandauLifshitz1}. 
The precession frequency is given by
\begin{align}
\Omega_p = \frac{\pi b}{2K(\tilde{m})}\left[\frac{(I_3-I_2)(I_3-I_1)}{I_1I_2} \right]^{1/2},
\end{align}
where $K(\tilde{m})$ is the complete elliptic integral of the first kind, with the ellipticity parameter $\tilde{m}$ given by
\begin{align}
\tilde{m} = \frac{(I_2-I_1)I_1a^2}{(I_3-I_2)I_3b^2}.
\end{align}
The rotation frequency is given by
\begin{align}\label{Eq: Omega_rot}
\Omega_\text{rot} &= \frac{\tilde{M}}{I_1} + \frac{2b}{K(\tilde{m})}\left[\frac{(I_3-I_2)(I_3-I_1)}{I_1I_2} \right]^{1/2} \nonumber \\ &\times \sum_{n=1}^{\infty} \frac{q^n}{1-q^{2n}}\sinh \left(2\pi n c_1\right)-\Omega_p,
\end{align}
where $q = \exp \left\{-\pi K(1-\tilde{m})/K(\tilde{m}) \right\}$,
and $c_1$ satisfies the following equation
\begin{equation}
\text{sn} \left[2ic_1K(\tilde{m}), \tilde{m}\right] = i\frac{I_3 b}{I_1 a}.
\end{equation}
It is evident from the set of Equations \eqref{Eq: GW polarization} that for a tri-axial system, GW is associated with three 
frequencies, which implies that one should expect three distinct lines in the spectrum. In reality, not only three in the spectrum, but 
lines with higher frequencies, arising from higher order terms, may also be present, whose, however, the intensity is suppressed \citep{Maggiore}.


\subsection{GW due to breaking of axial symmetry through rotation}

One more possibility of generating GW from an isolated WD is by breaking the pre-existence axial symmetry through rotation. A WD can be 
axially symmetric if it possesses a magnetic field (either toroidal or poloidal or any other suitable mixed field configuration). Now, 
if the WD rotates with a misalignment between its rotation and magnetic axes (similar configuration like a neutron star pulsar), it can 
emit gravitational radiation continuously. For such an object, with $\chi$ being the angle between the magnetic field and rotation axes, 
using Equation \eqref{Eq: Linearized tensorial amplitude}, the two tensorial polarizations of GW are given by \citep{1979PhRvD..20..351Z,1996A&A...312..675B}
\begin{equation}\label{Eq: GW polarization_magnetic}
\begin{aligned}
h_+ &= \tilde{A}_{+,1}\cos\left({\Omega_\text{rot} t}\right) - \tilde{A}_{+,2}\cos\left({2\Omega_\text{rot} t}\right), \\
h_\times &= \tilde{A}_{\times,1}\sin\left({\Omega_\text{rot} t}\right) - \tilde{A}_{\times,2}\sin\left({2\Omega_\text{rot} t}\right),
\end{aligned}
\end{equation}
where
\begin{equation}\label{Eq: GW polarization_magnetic amplitude}
\begin{aligned}
\tilde{A}_{+,1} &= \tilde{h}_0 \sin 2\chi \sin i \cos i, \\
\tilde{A}_{+,2} &= 2\tilde{h}_0 \sin^2\chi (1 + \cos^2 i), \\
\tilde{A}_{\times,1} &= \tilde{h}_0 \sin 2\chi \sin i, \\
\tilde{A}_{\times,2} &= 4\tilde{h}_0 \sin^2\chi \cos i,
\end{aligned}
\end{equation}
with 
\begin{equation}\label{Eq: h0 magnetic}
\tilde{h}_0 = \frac{G}{c^4}\frac{\Omega_\text{rot}^2(I_3-I_1)}{r}.
\end{equation}
Hence, in this configuration, continuous GW is emitted at two frequencies viz. $\Omega_\text{rot}$ and $2\Omega_\text{rot}$.


\section{Strength of GW emitted from an isolated WD} \label{Strength of GW emitted from an isolated WD}

\begin{figure}[!htpb]
	\centering
	\includegraphics[scale=0.5]{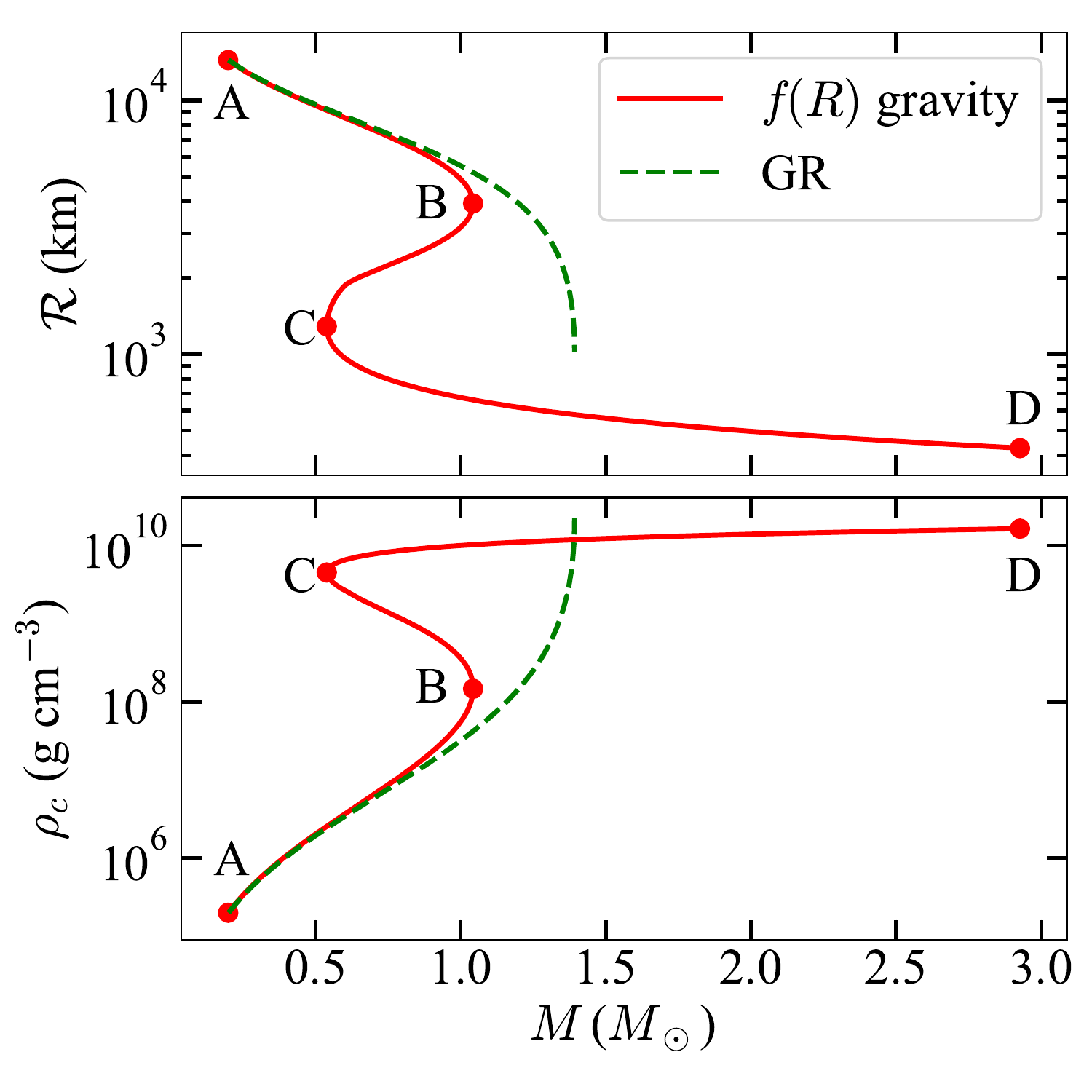}
	\caption{The variation of radius and central density with respect to the mass of the WD.}
	\label{Fig: Mass_radius plot}
\end{figure}
In this section, we discuss the strength of GW emitted from an $f(R)$ gravity induced isolated WD. 
We earlier showed that in the presence of $f(R)$ gravity with $f(R) = R+\alpha R^2(1-\gamma R)$, 
where $\alpha = 3\times 10^{14}$ cm$^2$ and $\gamma = 4 \times 10^{16}$ cm$^2$, it is possible 
to obtain sub-Chandrasekhar limiting mass WDs as well as super-Chandrasekhar WDs, along with the 
regular WDs just varying the central density $\rho_c$ of the WDs \citep{2018JCAP...09..007K}. 
We also showed that, with the chosen values of the parameters, this model is valid in terms of 
the solar system test \citep{2014IJMPD..2350036G}. For demonstration, we recall the key 
results from that work, depicted in Figure \ref{Fig: Mass_radius plot}, which shows the variation 
of radius $\mathcal{R}$ and $\rho_c$ with respect to $M$. WDs following GR are shown in green 
dashed line and the red solid line corresponds to the $f(R)$ gravity induced WDs.
Indeed, there are some WDs, e.g., EG 50, GD 140, J2056-3014, etc. \citep{1998ApJ...494..759P,2020ApJ...898L..40L},
which do not follow the standard Chandrasekhar mass--radius relation.
Moreover, it is to be noted that we adopt the perturbative calculations 
and consider the exterior solution of the WD to be the Schwarzschild solution 
while obtaining the mass--radius curve in $f(R)$ gravity. As a result, $R$
is asymptotically flat outside the WD
\citep{2014PhRvD..89f4019G,2016PhRvD..93b3501C}.
Moreover, in perturbative analysis, a WD is unstable under the radial perturbation if it falls in the 
branch where $\partial M/ \partial \rho_c <0$ ($\partial M/ \partial \rho_c >0$ is known as 
the positivity condition, which is a necessary condition for stability) \citep{compact,glendenning2010}. 
In perturbative method, while obtaining the mass--radius curve, $\partial M/ \partial \rho_c>0$ 
still holds good to define the stability condition like GR, because any additional terms will be just 
perturbative corrections to the zeroth-order quantity which are usually small.
From the figure, we observe that at low $\rho_c$, the effect of modified gravity is negligible, as both 
the curves overlap with each other (mostly in the branch AB). As $\rho_c$ increases, 
the mass--radius curve reaches a maximum at the mass $\sim 1 M_\odot$ and 
$\rho_c \sim 1.5 \times 10^8$ g cm$^{-3}$ (point B). Beyond this $\rho_c$, the curve turns back which 
violates the positivity condition and, hence, BC is an unstable branch. Therefore, point B 
corresponds to the sub-Chandrasekhar limiting mass WD. Further, reaching a minimum value, 
the curve again turns back from the point C, and quickly enters in the super-Chandrasekhar 
WD regime following $\partial M/ \partial \rho_c >0$. Since the branch CD is stable, the 
super-Chandrasekhar WDs are stable under radial perturbation. The maximum $\rho_c$
is chosen in such a way that it does not violate any of the known physics for CO WDs, such as 
neutron drip \citep{compact}, pycno-nuclear reactions and inverse beta decay \citep{2019ApJ...879...46O}, etc.
The empirical relation between 
$\mathcal{R}$ and $M$, in the branch AB is approximately $\mathcal{R}\propto M^{-2/5}$, in 
the branch BC is $\mathcal{R}\propto M$, and that of in the branch CD is $\mathcal{R}\propto M^{-1/2}$. 
In this way, this form of $f(R)$ gravity, with the chosen parameters, can explain both the sub- and 
super-Chandrasekhar limiting mass WDs just by varying the $\rho_c$. It is however evident that there 
is no super-Chandrasekhar mass-limit in this model. Such a mass-limit is possible if we consider 
higher-order corrections (at least $16^\mathrm{th}$ order) to the Starobinsky model as discussed earlier \citep{2018JCAP...09..007K}.
The beauty of this model is that at the low-density limit, 
the extra terms of modified gravity have a negligible effect, and 
GR is enough to describe the underlying physics. In the case of a star, its density is relatively small 
compared to the WD's density and, hence, even if the same $\alpha$ and $\gamma$ remain intact, their effect is 
not prominent in the star phase. It brings in significant effect only when the
star becomes a WD depending on the density.
Now, based on this mass--radius relation, we discuss the corresponding 
strength of GW separately for each of 
the possibilities, mentioned in the previous section.


\subsection{Presence of roughness at WD's surface}

\begin{figure}[!htpb]
	\centering
	\includegraphics[scale=0.5]{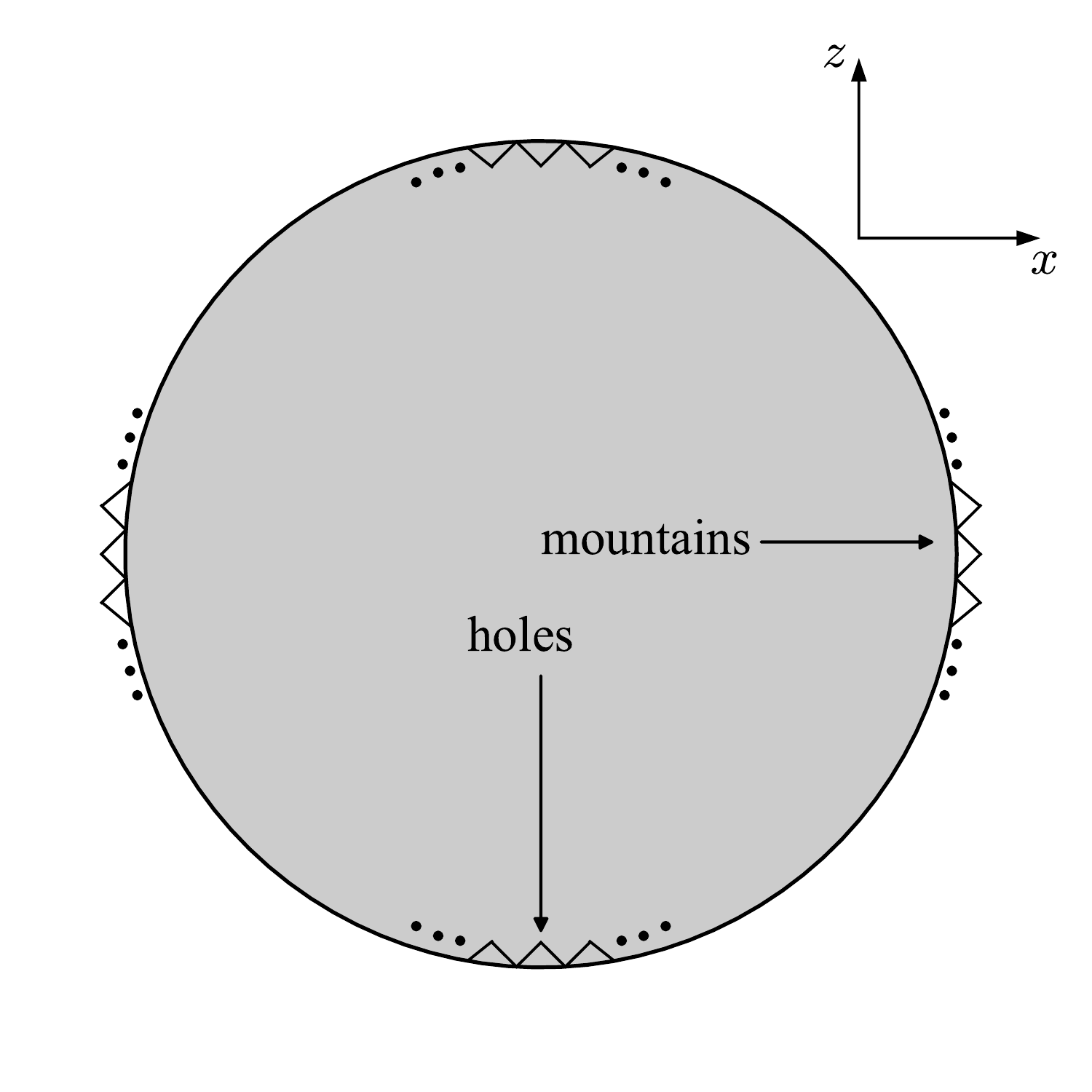}
	\caption{Exaggerated figure of the presence of mountains and holes at the surface of a WD.}
	\label{Fig: Mountain in WD}
\end{figure}

We have already mentioned that if a pre-existed tri-axial WD rotates, it can produce continuous gravitational radiation. One possibility 
for a WD of being tri-axial is due to the asymmetry of matter present at its surface. One can imagine this configuration as a similar 
structure of the earth or moon, where there are mountains and craters (holes) at the surface. We assume that there are excess 
`mountains' along the $x-$axis and `holes' along the $z-$axis of a WD, as shown in Figure \ref{Fig: Mountain in WD}. Such a 
configuration guarantees a tri-axial system with $I_1 < I_2 < I_3$. For simplicity, we assume $\Omega_1=\Omega_2=\Omega_3$.

\begin{figure}[!htpb]
	\centering
	\includegraphics[scale=0.5]{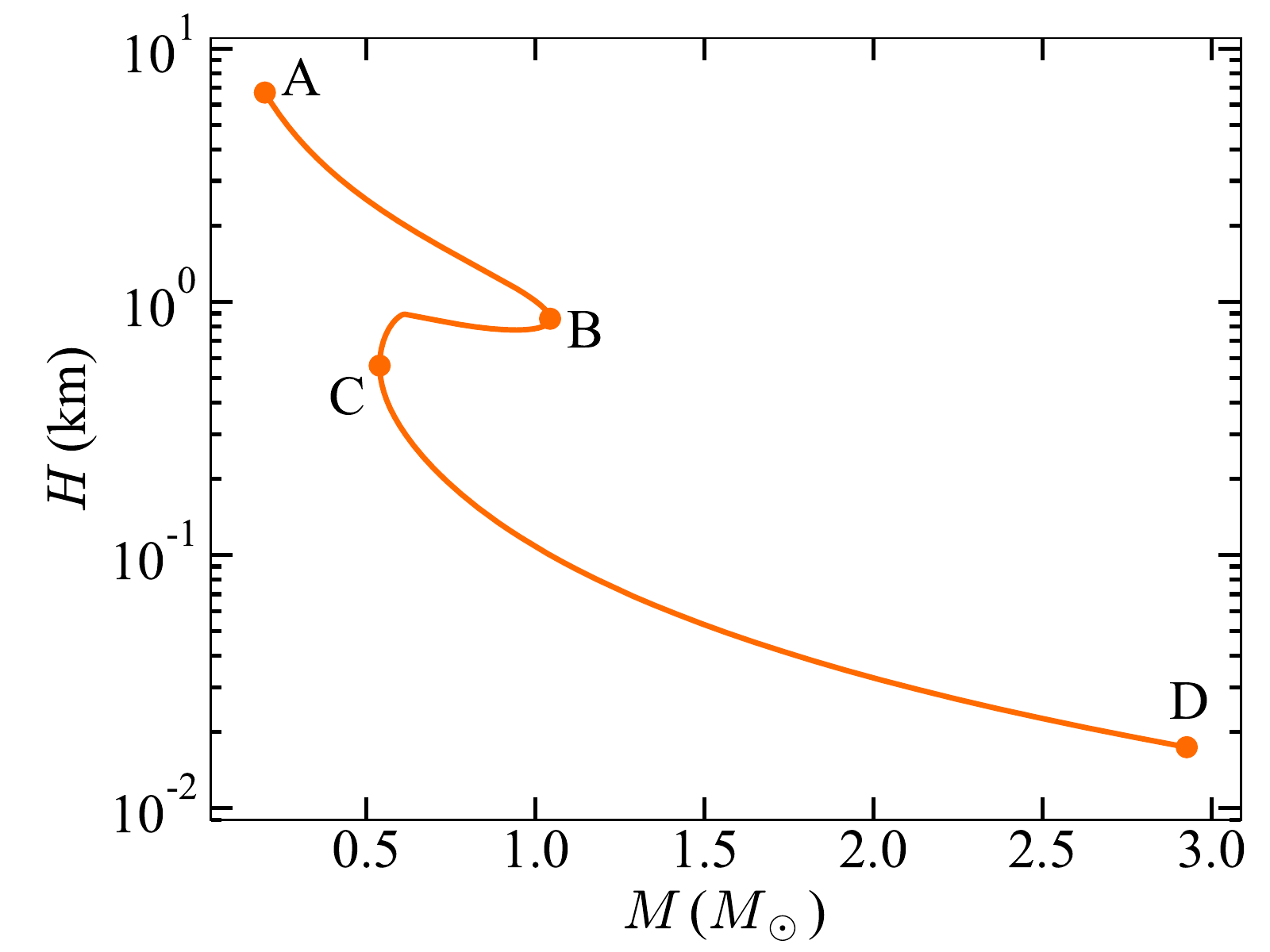}
	\caption{The maximum height of a mountain present on a WD as a function of mass.}
	\label{Fig: Maximum mountain height}
\end{figure}

Let us now check the maximum possible height of a mountain on a WD's surface. Assuming the mountains to be generated due to the shear in 
the outer envelope of WDs (same as the case for the earth where there are mountains at its crust), the maximum height of a mountain is 
given by \citep{2005Ap.....48...53S}
\begin{equation}
H = \frac{S}{\rho g},
\end{equation}
where $\rho$ is the average density of the mountain, $g$ the acceleration due to gravity at the surface of the WD, and $S$ the 
shear modulus which is given by \citep{mott1958theory,1971AnPhy..66..816B}
\begin{equation}
S = 0.295 ~Z^2 e^2 n_e^{4/3},
\end{equation}
with $e$ being the charge of an electron, $Z$ the atomic number, and $n_e$ the electron number density in the mountain. 
Substituting this expression, the maximum height of the mountain reduces to
\begin{equation}
H = 3.637 \times 10^{12} \frac{Z^2}{g} \rho^{1/3} ~\mathrm{cm}.
\end{equation}
Assuming the C-O WDs with the surface mostly consisting of He, we choose $Z=2$. Moreover, assuming the mountains' average density to be 
the density just below the surface of the WDs, i.e. $\rho \approx 10^{-6}\rho_c$, we obtain $H$ for various WDs with different values of 
$\rho_c$, which is shown in Figure \ref{Fig: Maximum mountain height}. From the empirical relations mentioned in the previous section, 
we also obtain the same between $H$ and $M$: in the branch AB, $H\propto M^{-2/5}$, in CD, $H\propto M^{-7/6}$, while in BC, $H$ is 
nearly constant.

\begin{figure*}[!htbp]
	\centering
	\subfigure[]{\includegraphics[scale=0.27]{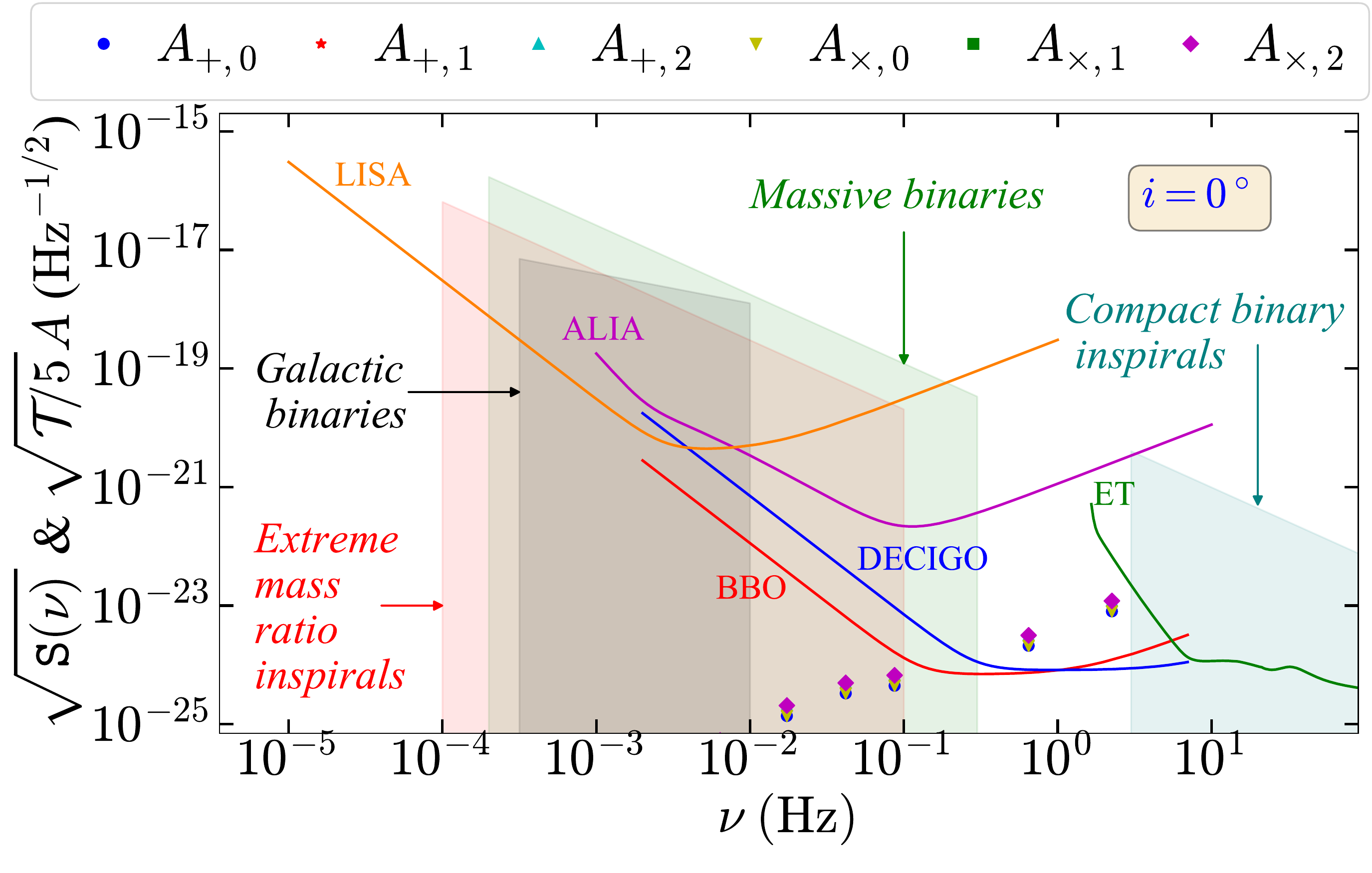}}
	\subfigure[]{\includegraphics[scale=0.27]{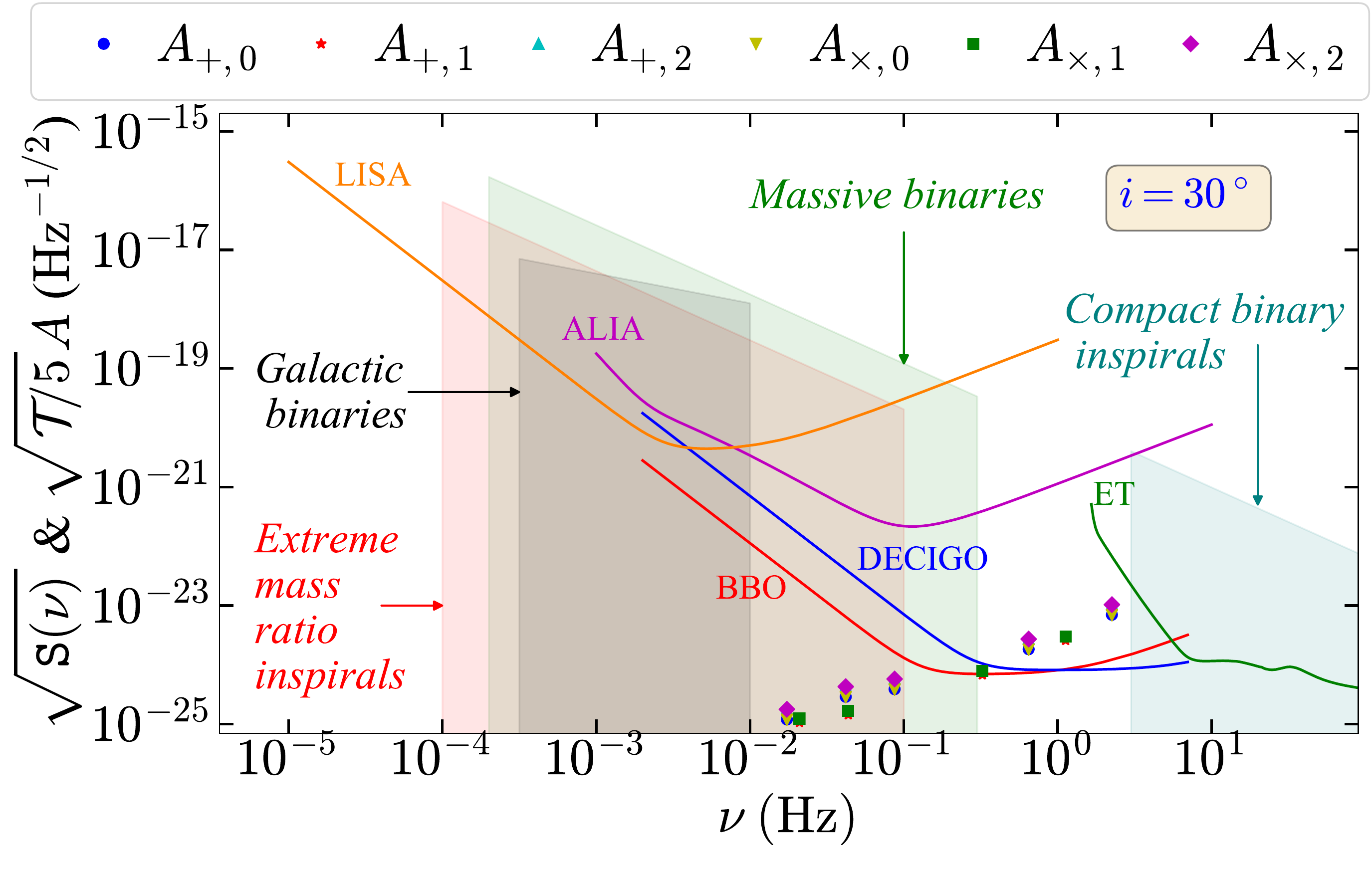}}
	\subfigure[]{\includegraphics[scale=0.27]{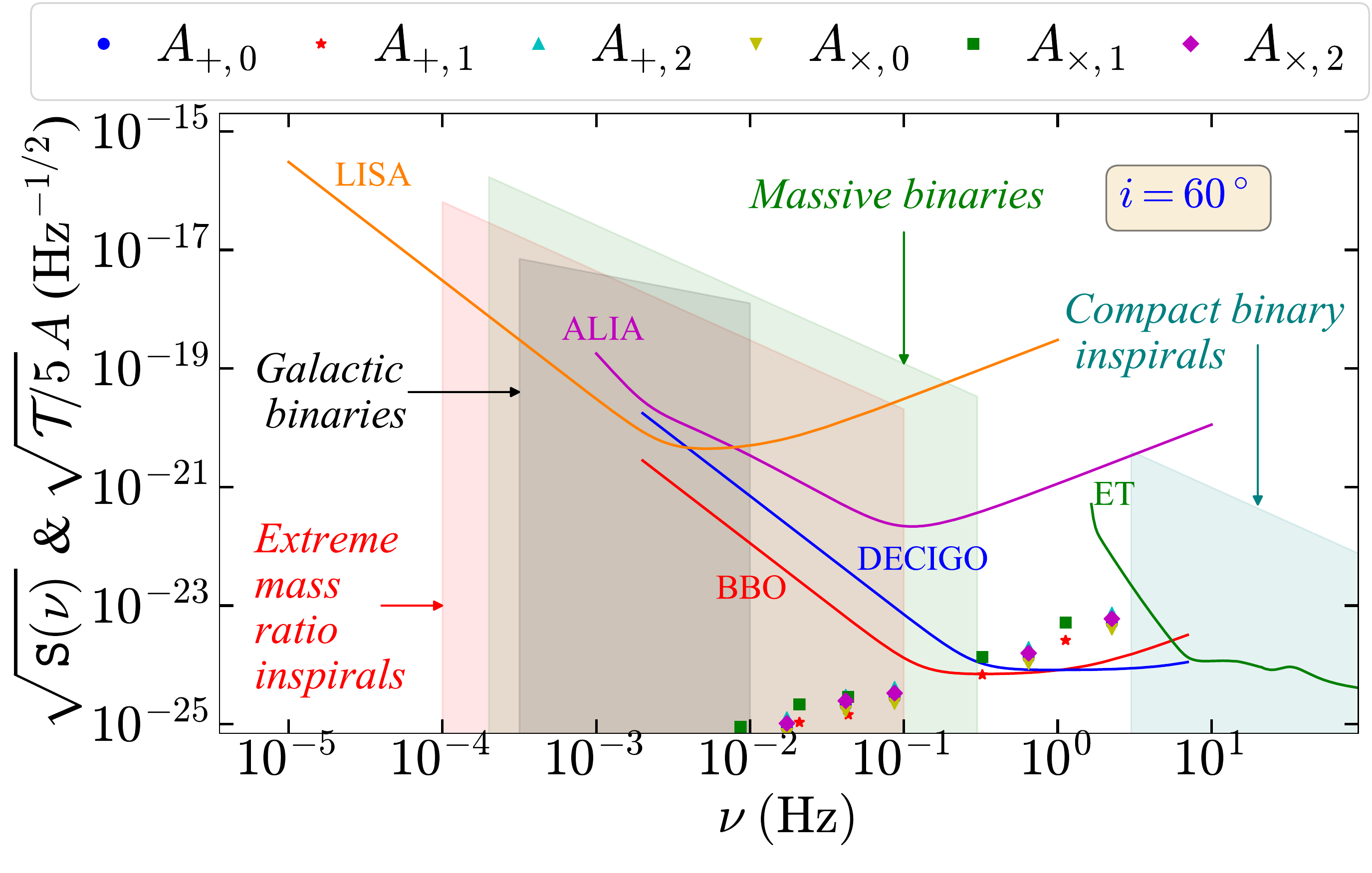}}
	\subfigure[]{\includegraphics[scale=0.27]{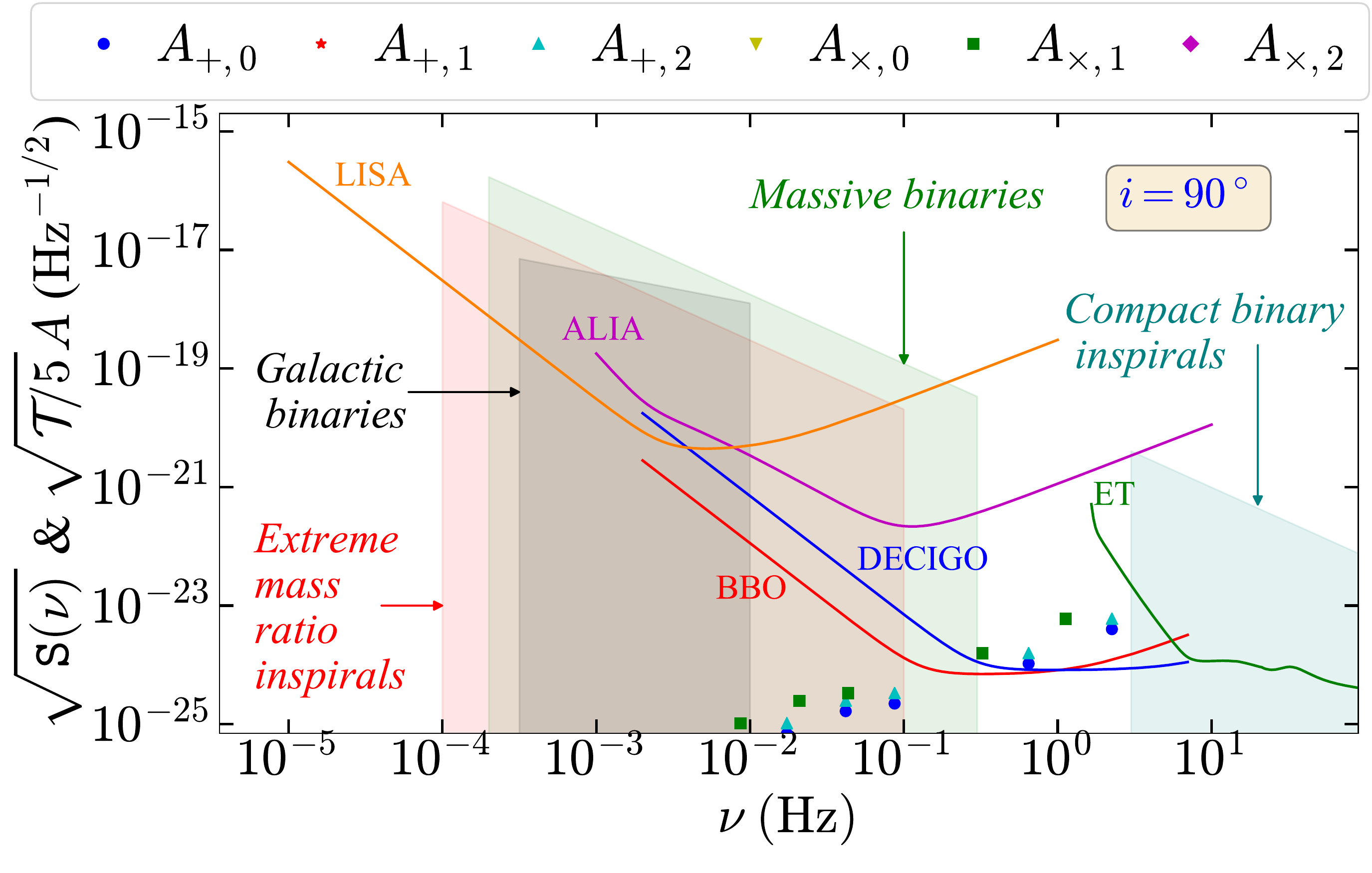}}
	\caption{$\sqrt{\mathcal{T}/5}A$ for $f(R)$ gravity induced WDs with rough surfaces for different $i$ over 5 s integration time along with various detectors' PSD.}
	\label{Fig: Detectors_mountain}
\end{figure*}
\begin{figure*}[!htbp]
	\centering
	\subfigure[~WDs with surface roughness.]{\includegraphics[scale=0.27]{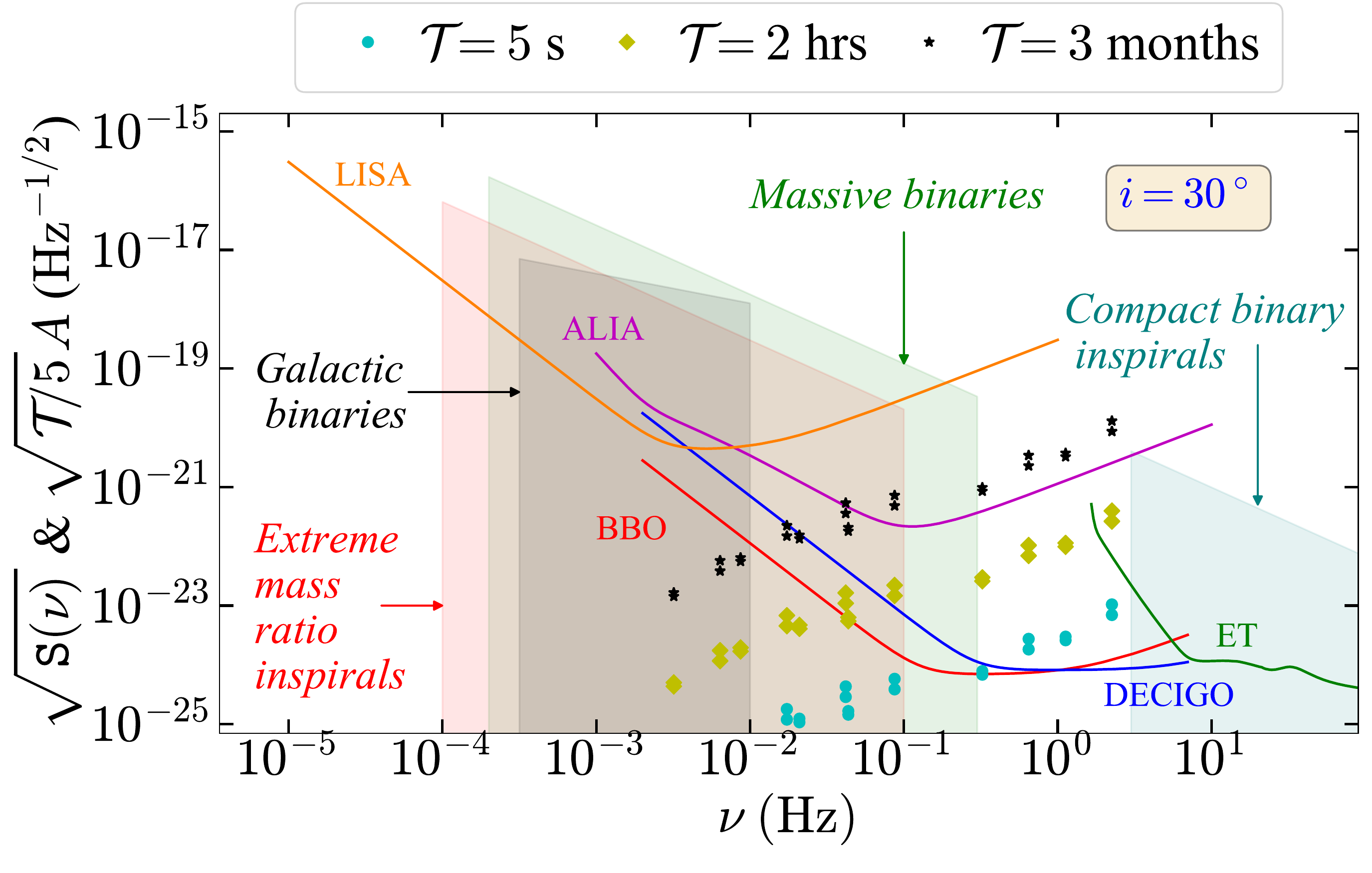}}
	\subfigure[~WDs with magnetic field.]{\includegraphics[scale=0.27]{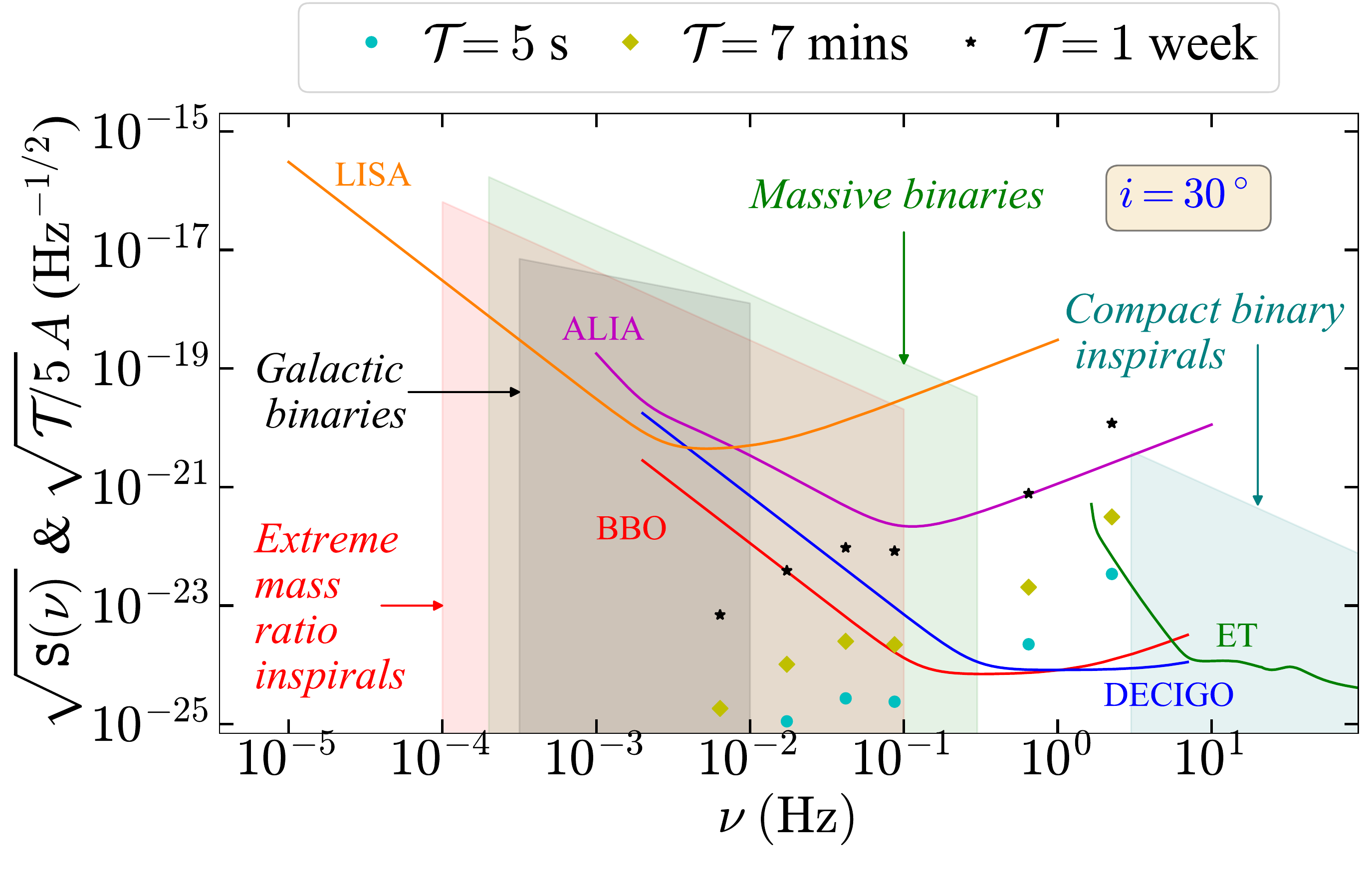}}
	\caption{$\sqrt{\mathcal{T}/5}A$ for $f(R)$ gravity induced WDs with $i=30\degree$ and different integration time along with various detectors' PSD.}
	\label{Fig: Compare integration time}
\end{figure*}
\begin{figure}[!htpb]
	\centering
	\includegraphics[scale=0.5]{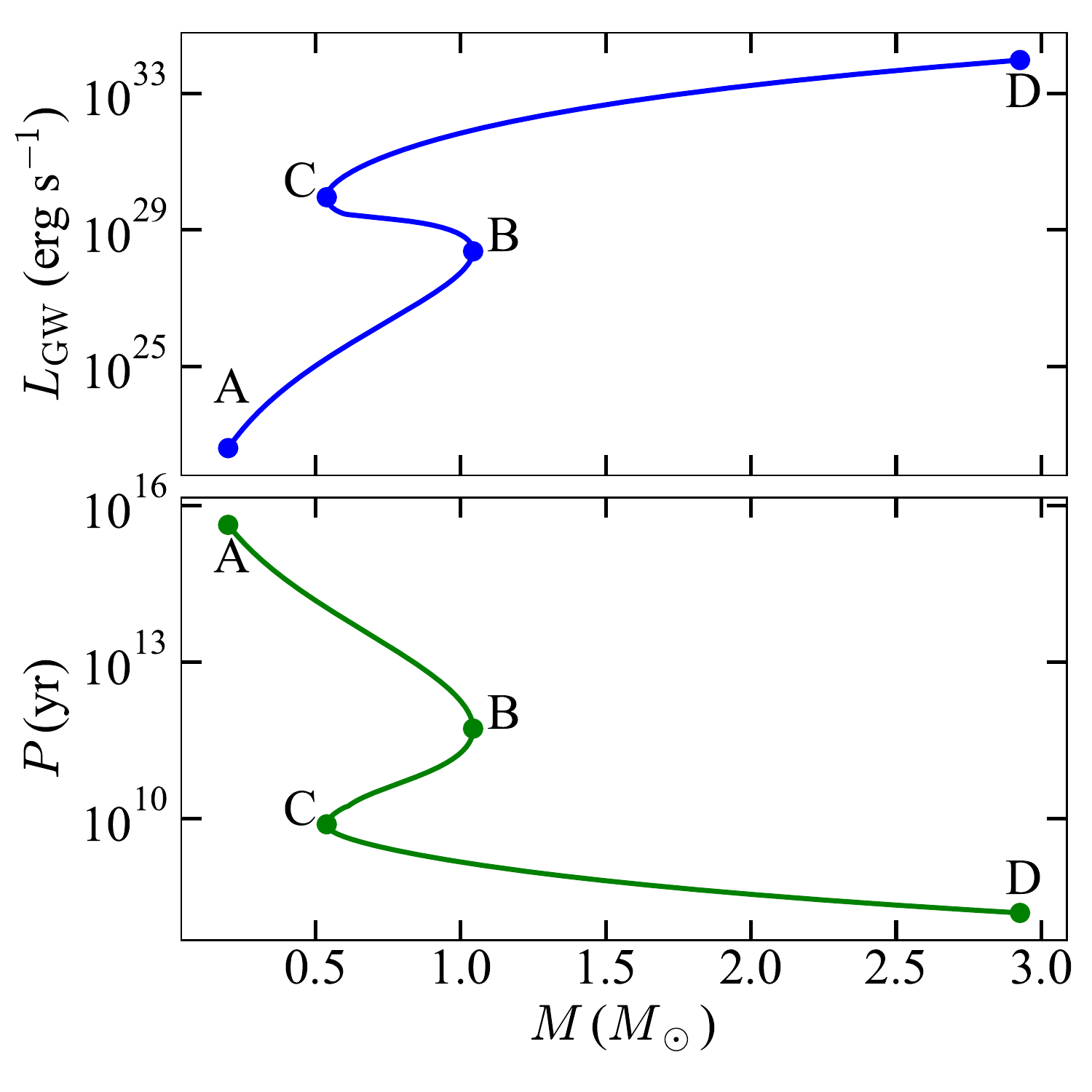}
	\caption{The variation of $L_\text{GW}$ and $P$ with respect to the mass of WD with mountains and holes.}
	\label{Fig: Luminosity mountain}
\end{figure}

It is evident from Figure \ref{Fig: Maximum mountain height} that $H \ll\mathcal{R}$. However, if there is a series of mountains of similar 
heights on the $x-$direction and big holes on the $z-$direction, as shown in Figure \ref{Fig: Mountain in WD}, effective radii of 
the WDs alter. The effective radii in $x-$, $y-$ and $z-$directions become $\mathcal{R}+H$, $\mathcal{R}$ and $\mathcal{R}-H$ respectively, 
resulting in a tri-axial system. Hence, the moments of inertia along different directions are given by
\begin{equation}
\begin{aligned}
I_1 &= \frac{M}{5}\left[\mathcal{R}^2 + \left(\mathcal{R}-H\right)^2\right], \\ I_2 &= \frac{M}{5}\left[\left(\mathcal{R}+H\right)^2 + \left(\mathcal{R}-H\right)^2\right], \\ I_3 &= \frac{M}{5}\left[\mathcal{R}^2 + \left(\mathcal{R}+H\right)^2\right].
\end{aligned}
\end{equation}
Moreover, we know that rotation can also increase the mass and radius of a WD. Hence, we assume the angular 
frequency to be $1/10$th of the maximum possible angular frequency, i.e., $\Omega = 1/10\sqrt{GM/\mathcal{R}^3}$, such that it does not 
affect the mass and size of the WD. Using the set of Equations \eqref{Eq: GW polarization amplitude}, we obtain the dimensionless 
strain amplitude of GW, $A$ (e.g., $A_{+,0}$, $A_{\times,0}$, etc.), emitted from WDs with rough surface, corresponding to all three 
frequencies of the spectrum. Moreover, we know that the integrated signal to noise ratio (SNR) increases if we observe the source for 
a long period of time $\mathcal{T}$. The relation between SNR and $\mathcal{T}$ is given by \citep{Maggiore,2013PhRvD..88l4032T,2019Univ....5..217S}
\begin{equation}
\mathrm{SNR} = \frac{1}{\sqrt{5}} \left(\frac{\mathcal{T}}{\mathtt{S}(\nu)}\right)^{1/2}A,
\end{equation}
where $\mathtt{S}(\nu)$ is the power spectral density (PSD) at a frequency $\nu$. We show PSD of various detectors as a function of 
frequency\footnote{\url{http://gwplotter.com/}} \citep{2009LRR....12....2S,2015CQGra..32a5014M} in Figure \ref{Fig: Detectors_mountain}, 
along with $\sqrt{\mathcal{T}/5}A$ for $f(R)$ gravity induced WDs with $\rho_c = 10^6, 10^7, 10^8, 10^9, 10^{10}, 1.66\times10^{10}$ 
g cm$^{-3}$ over $\mathcal{T}=5$ s assuming $r=100$ pc. It is evident that many of these dense WDs will readily, or at most in a few seconds, 
be detected by DECIGO and BBO with $\mathrm{SNR} \gtrsim 5$. Since the signal is continuously emitted for a long duration, the significantly 
super-Chandrasekhar WDs, being smaller in size, can also be detected by the Einstein Telescope with $\mathrm{SNR} \approx 5$ if the signal is 
integrated over $\mathcal{T}\sim 102$ mins. However, to detect these WDs by ALIA or LISA with the same SNR, the integration time turns out to 
be $\mathcal{T} \sim 3$ months and $\mathcal{T} \sim 40000$ yrs respectively. Hence, it seems impossible for LISA to detect the GW signal 
from WDs with rough surfaces. Figure \ref{Fig: Compare integration time}(a) shows $\sqrt{\mathcal{T}/5}A$ for different integration times for
these WDs. It is evident from this figure that SNR increases as the integration time increases, allowing possibility of detecting these WDs even by the 
Einstein Telescope and ALIA.
Furthermore, due to the emission of gravitational radiation, it is associated with the quadrupolar luminosity, given by \citep{1980PhRvD..21..891Z}
\begin{align}
L_\text{GW} &= -\dv{E}{t} = - I_3 \Omega_\text{rot} \dot{\Omega}_\text{rot} \nonumber \\ 
&\approx \frac{32G}{5c^5}b^6 \left(I_2-I_1\right)^2 + \frac{2G}{5c^5}a^2b^4\left(I_3-\frac{I_1+I_2}{2}\right)^2.
\end{align}
Since there is no magnetic field in this WD configuration, there will be no associated electromagnetic counterpart. 
In other words, these WDs do not emit any dipole radiation. Nevertheless, due to the emission of GW radiation, the WD starts spinning down, 
i.e., $\Omega_\text{rot}$ decreases with time. After a certain period 
of time (characteristic timescale, $P$, of a WD pulsar), it will lose all its rotational energy and can no longer radiate any 
gravitational radiation. Using the expression for $L_\text{GW}$, we obtain
\begin{align}
P \approx \frac{135 I_3 c^5}{2 G \Omega_\text{rot}^4} \frac{1}{64 Y^3 \left(I_2-I_1\right)^2 + XY^2 \left(2I_3-I_1-I_2\right)^2},
\end{align}
with
\begin{align*}
X = 1+\frac{I_2(I_3-I_2)}{I_1(I_3-I_1)} ~~\mathrm{and}~~ Y = 1+\frac{I_2(I_2-I_1)}{I_3(I_3-I_1)}.
\end{align*}
Figure \ref{Fig: Luminosity mountain} shows the variation of $L_\text{GW}$ and $P$ with respect to $M$ of WD. 
The empirical relations for $L_\text{GW}$ and $P$ in various branches are given in Table \ref{Table: empirical relations}. It is evident 
that the life-time of massive WD pulsars is shorter than that of the lighter WDs.

\begin{table*}[!htpb]
	\caption{The empirical relations of various quantities with respect to mass of the WDs.}
	\label{Table: empirical relations}
	\centering
	\begin{tabular}{|l|l|l|l|}
		\hline
		Quantity & AB branch & BC branch & CD branch \\
		\hline
		Radius & $\mathcal{R} \propto M^{-2/5}$ & $\mathcal{R} \propto M$ & $\mathcal{R} \propto M^{-1/2}$\\
		Maximum height & $H \propto M^{-2/5}$ & $H \approx$ constant & $H \propto M^{-7/6}$\\
		Luminosity & $L_\text{GW} \propto M^{3}$ & $L_\text{GW} \approx$ constant & $L_\text{GW} \propto M^{5/2}$\\
		Timescale & $P \propto M^{-23/5}$ & $P \propto M$ & $P \propto M^{-5}$\\
		\hline
	\end{tabular}
\end{table*}


\subsection{Presence of magnetic field in WD}

\begin{figure}[!htpb]
	\centering
	\includegraphics[scale=0.5]{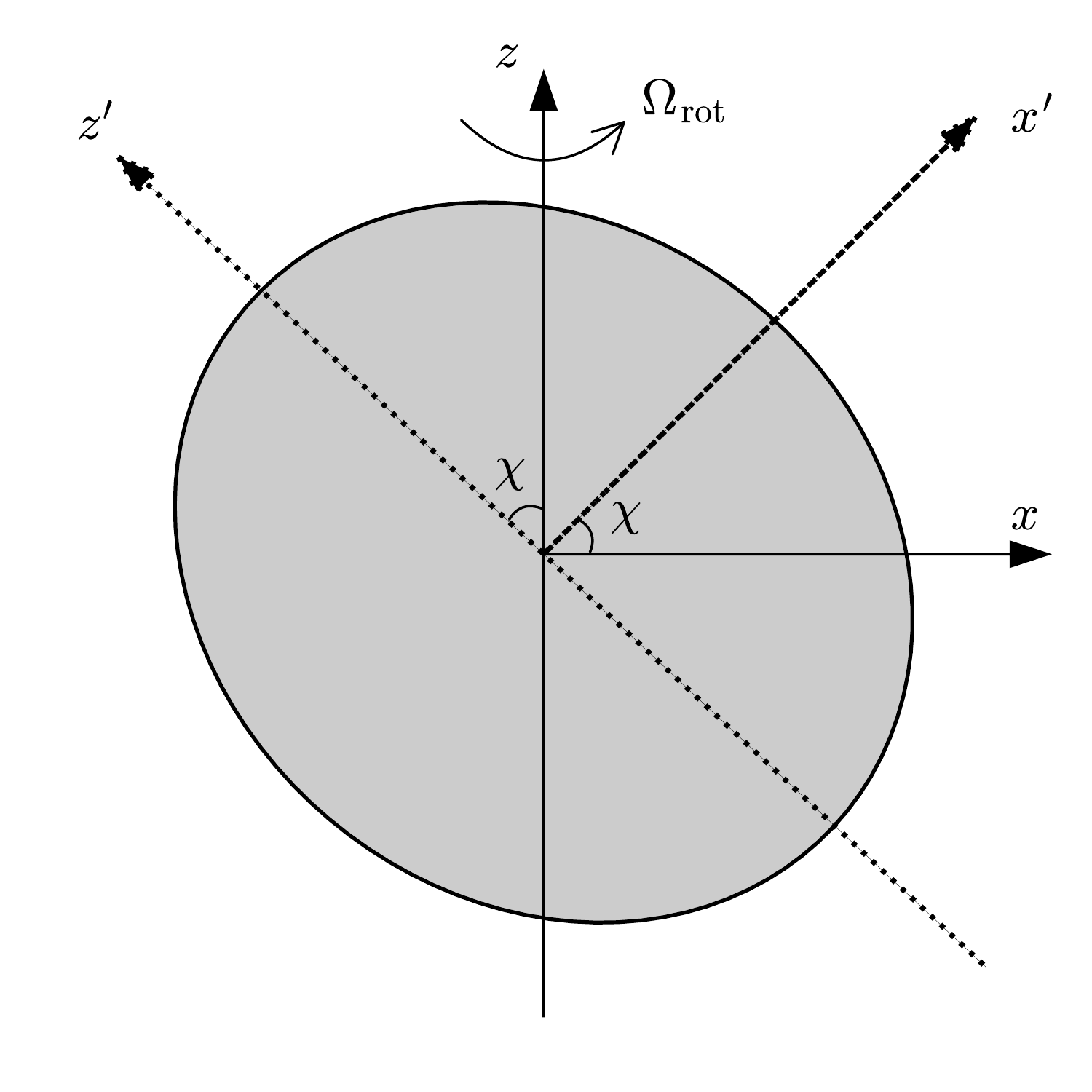}
	\caption{Cartoon diagram of a magnetized WD with magnetic field is along $z'-$axis and rotation is along $z-$axis.}
	\label{Fig: Magnetized WD}
\end{figure}

\begin{figure*}[!htbp]
	\centering
	\subfigure[]{\includegraphics[scale=0.27]{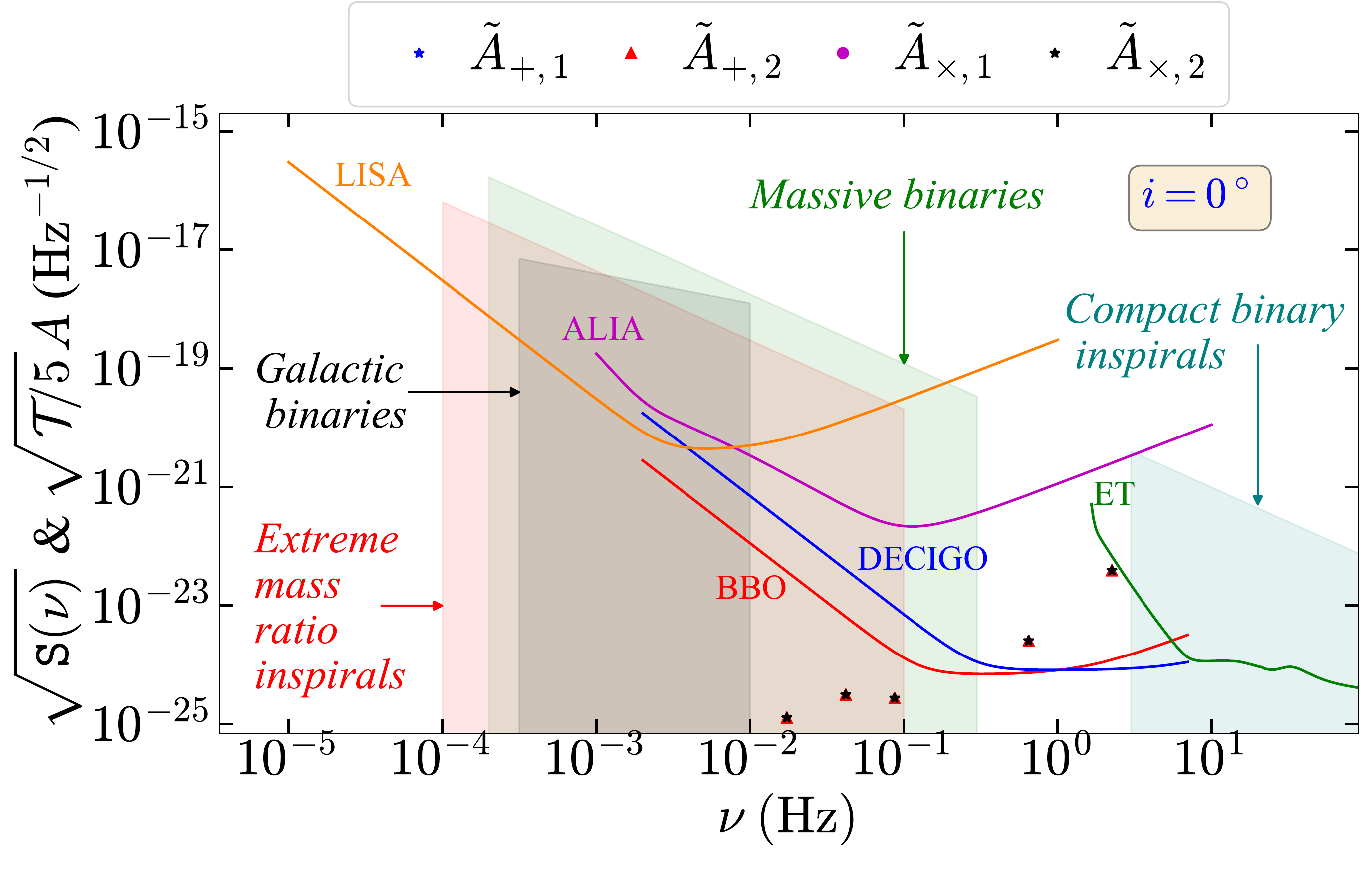}}
	\subfigure[]{\includegraphics[scale=0.27]{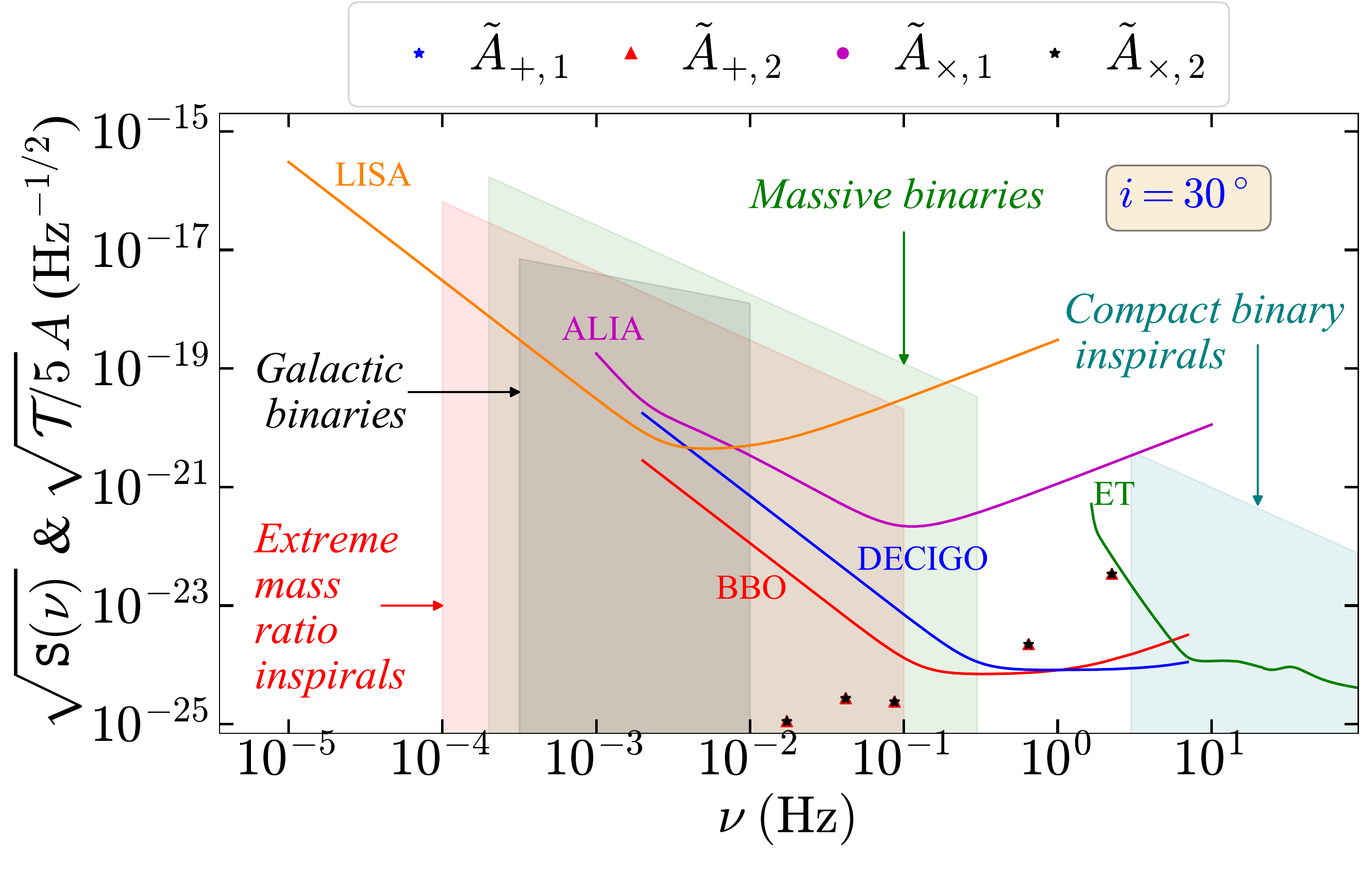}}
	\subfigure[]{\includegraphics[scale=0.27]{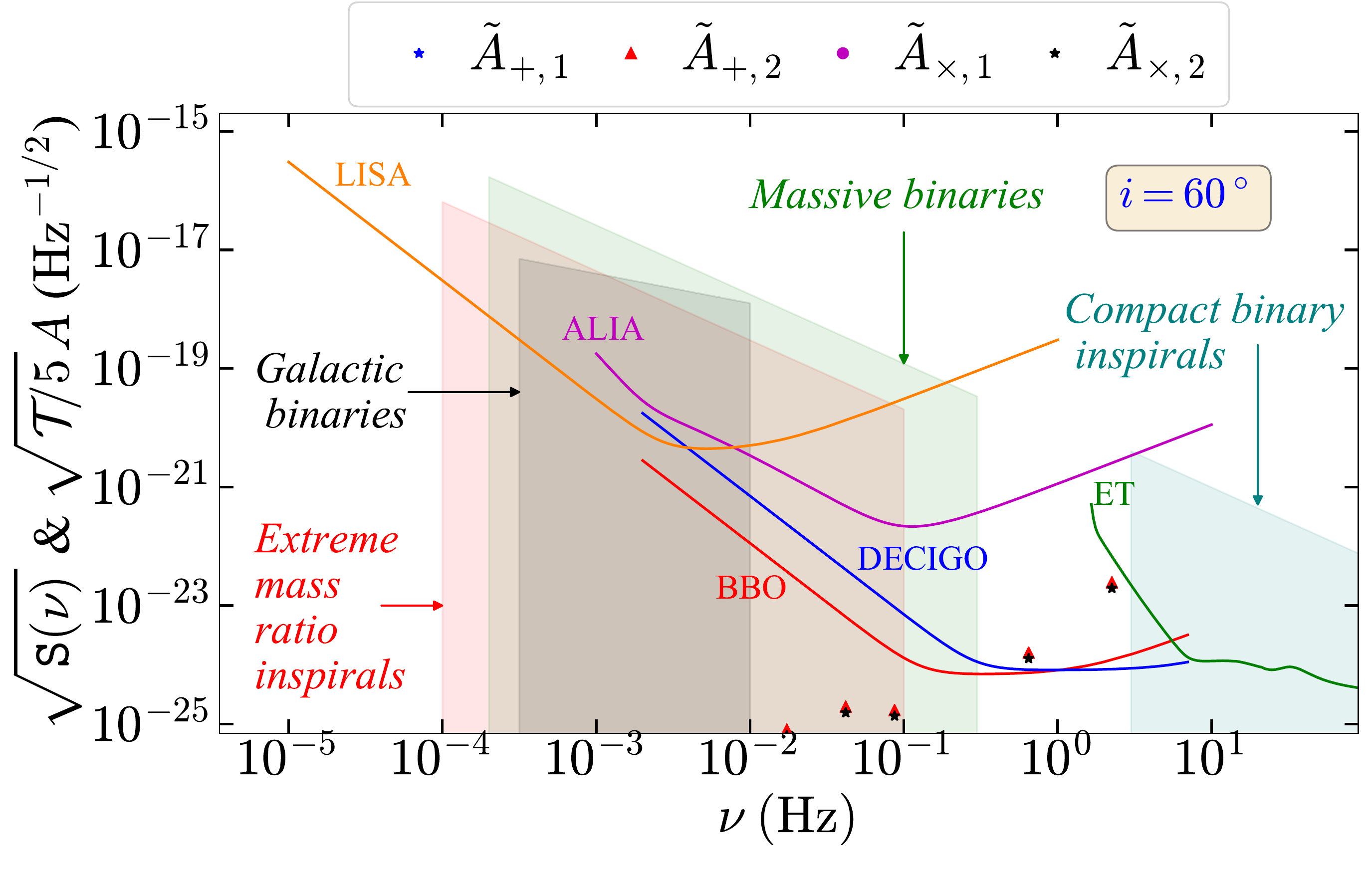}}
	\subfigure[]{\includegraphics[scale=0.27]{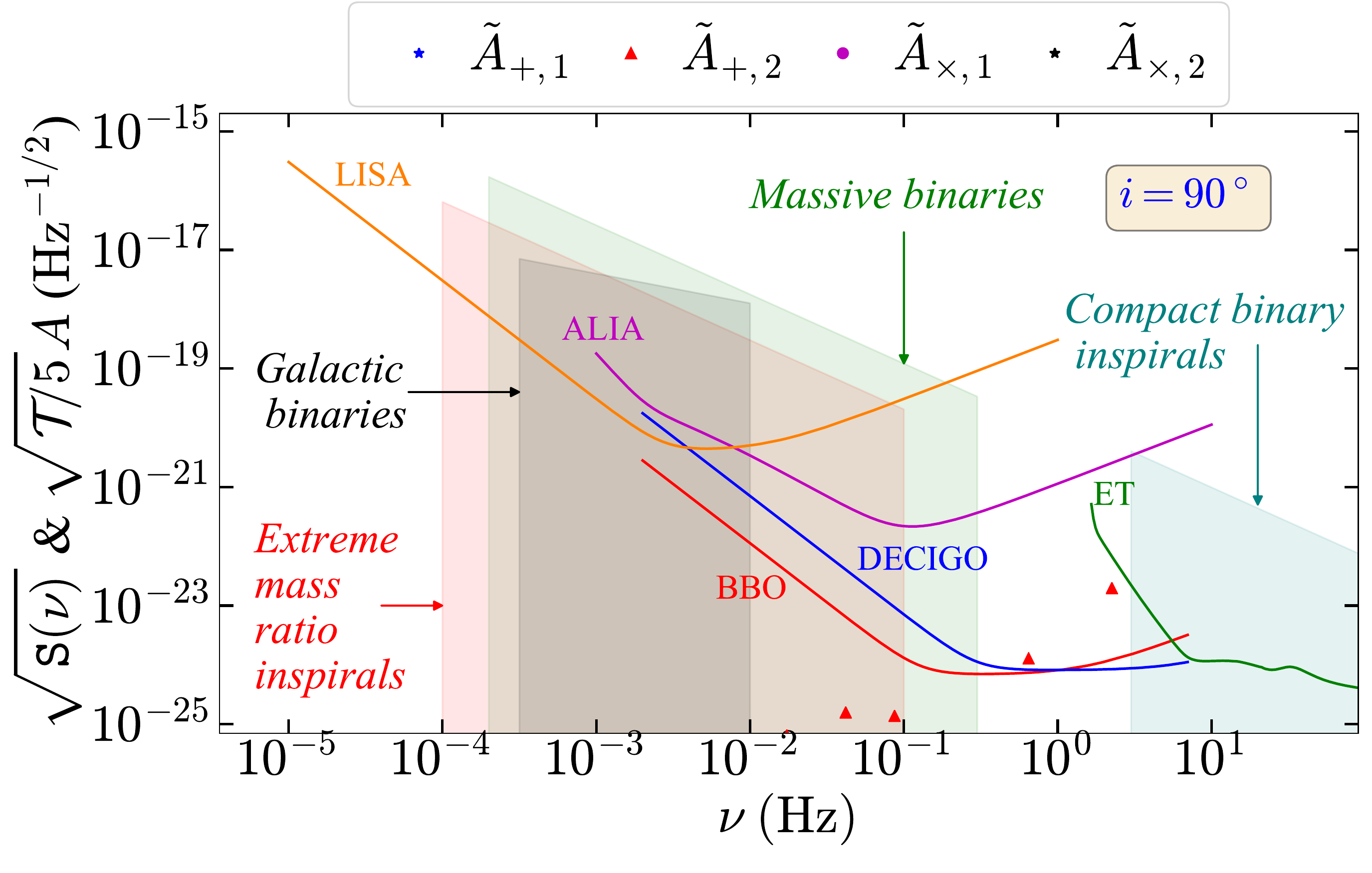}}
	\caption{$\sqrt{\mathcal{T}/5}A$ for $f(R)$ gravity induced weakly magnetized WD pulsars for different $i$ over 5 s integration time along with various detectors' PSD.}
	\label{Fig: Detectors_magnetic}
\end{figure*}

\begin{figure}[!htpb]
	\centering
	\includegraphics[scale=0.5]{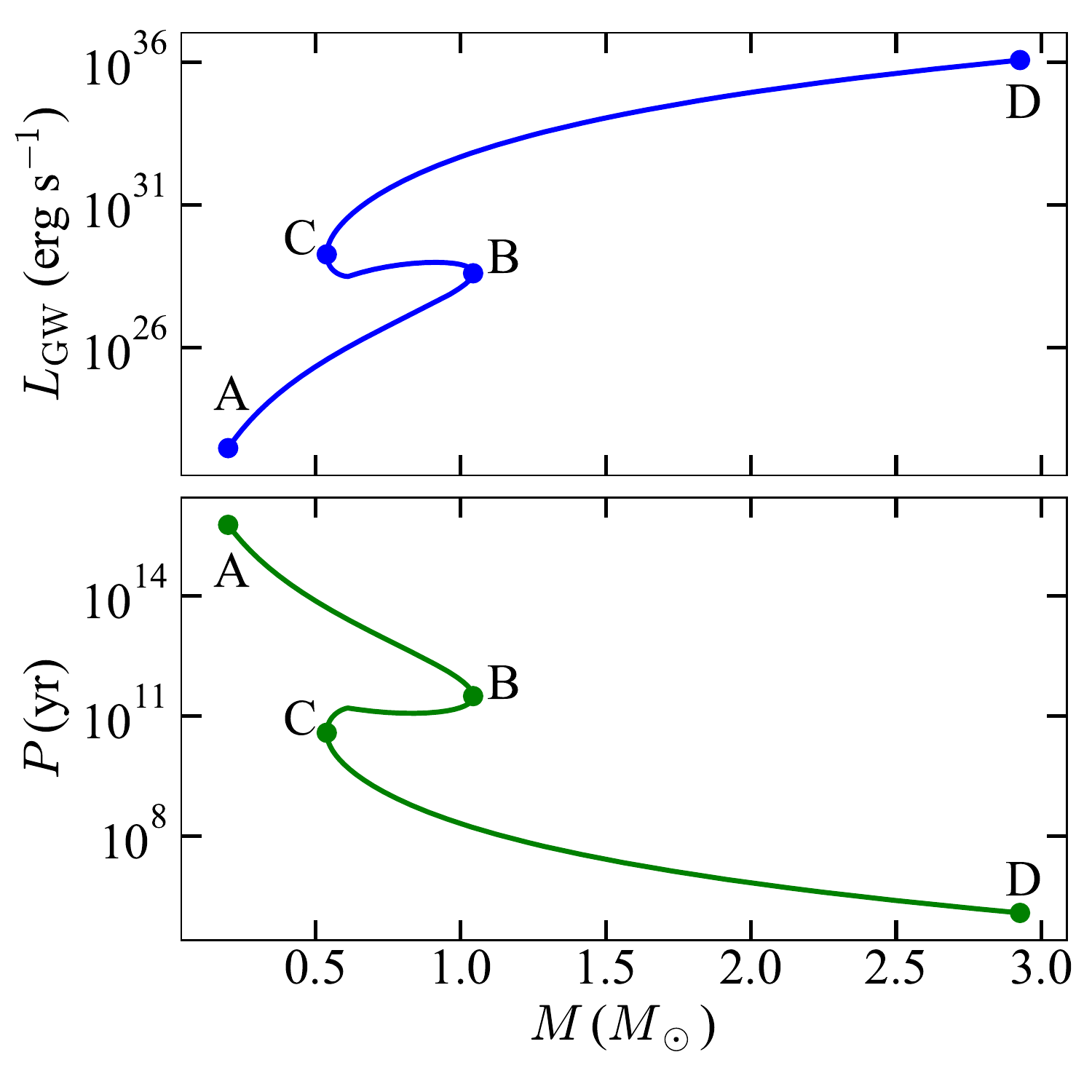}
	\caption{Same as Figure \ref{Fig: Luminosity mountain} except that the WDs have magnetic field instead of mountains.}
	\label{Fig: Luminosity magnetic}
\end{figure}

As mentioned in the previous section, if a magnetized WD rotates with a misalignment between its magnetic field and rotation axes 
(similar configuration of a pulsar), it can emit continuous GW. We already provided a detailed discussion on GW emitted from WDs 
with different magnetic field geometries and strengths in GR \citep{2019MNRAS.490.2692K,2020ApJ...896...69K}. Figure 
\ref{Fig: Magnetized WD} shows an illustrative diagram of a magnetized WD where the magnetic field is along $z'-$axis and rotation is 
along the $z-$axis, with $\chi$ being the angle between these two axes. We calculate the amplitude of GW using the set of Equations 
\eqref{Eq: GW polarization_magnetic amplitude} assuming the difference in radii of the WD between those along $x-$ and $z-$axes be 
$0.01 \%$, i.e., $\epsilon = \abs{I_3-I_1}/I_3 \approx 2\times10^{-4}$, due to the presence of a very weak magnetic field and slow rotation. 
The choice of weak fields and slow rotation assures that the underlying WD mass--radius solutions do not practically differ from the 
solutions based on the $f(R)$ gravity without magnetic fields and rotation. In future, we plan to check rigorously by solving the set of 
equations, if indeed such $\epsilon$ is possible in the presence of weak magnetic fields and rotation keeping the mass and radius 
practically intact. As we will show below, however, the chosen $\epsilon$ appears to be minimally required value to have any appreciable 
effect. Nevertheless, there are examples of weakly magnetized WD pulsars, which can be explained even in GR framework, e.g. AE Aquarii 
\citep{1987ApJ...323L.131B}, AR Scorpii \citep{2016Natur.537..374M}, where magnetic fields hardly affect their mass--radius relations. 
Figure \ref{Fig: Detectors_magnetic} shows PSD as a function of frequency for various detectors along with 
$\sqrt{\mathcal{T}/5}A$ over 5 s integration time for various $f(R)$ gravity induced WD pulsars with different $i$ assuming $\chi=90\degree$ 
and $r=100$ pc. It is evident that while DECIGO and BBO can immediately detect such weakly magnetized super-Chandrasekhar WDs, 
the Einstein Telescope can detect them in $\mathcal{T}\sim 6$ mins with $\mathrm{SNR} \approx 5$ (see Figure \ref{Fig: Compare integration time}(b)). However, for ALIA and LISA, the 
corresponding integration time respectively turns out to be $\mathcal{T} \sim 5$ days and $\mathcal{T} \sim 25000$ yrs\footnote{Note that 
even if the threshold SNR for detection increases slightly (say 5 to 20), many of these sources can still be detected in a few seconds to a 
few days of integration time depending on the type of the detectors.}. 
Hence, it is also possible to detect such weakly magnetized WDs using ALIA, 
whereas for LISA, it is quite impossible. Figure \ref{Fig: Compare integration time}(b) depicts $\sqrt{\mathcal{T}/5}A$ for these WDs with 
different integration times to show that SNR increases if the integration time increases so that various detectors can detect them 
eventually. For such a system, the GW luminosity is given by \citep{1979PhRvD..20..351Z}
\begin{align}
L_\text{GW} \approx \frac{2G}{5c^5} \left(I_3-I_1\right)^2 \Omega^6 \sin^2\chi \left(1+15\sin^2\chi\right).
\end{align}
It is expected that a source can emit electromagnetic radiation in the presence of a magnetic field, 
and it is the dipole radiation in the case of a WD pulsar. However,
because of the presence of a weak magnetic field, the dipole radiation emitted from such a WD is minimal, and the corresponding dipole 
luminosity is negligible as compared to $L_\text{GW}$. Hence the spin-down timescale is mostly governed by $L_\text{GW}$, given by \citep{2020ApJ...896...69K} 
\begin{align}
P \approx \left(\frac{5I_3 c^5}{8 G \left(I_3-I_1\right)^2 \Omega^4}\right)\frac{1}{\sin^2\chi \left(1+15\sin^2\chi\right)}.
\end{align}
Figure \ref{Fig: Luminosity magnetic} shows the variation of $L_\text{GW}$ and $P$ with respect to $M$ for various WDs with 
$\chi = 90\degree$. The maximum $L_\text{GW}$ in the case of a WD is $\sim 10^{37}$ erg s$^{-1}$.
The empirical relations of $L_\text{GW}$ and $P$, in various branches, are same as the previous case provided in Table 
\ref{Table: empirical relations}. It is also clear from the figure that the massive WD pulsars are short-lived as compared to the lighter ones.


\section{Discussion} \label{Discussion}

From Figures \ref{Fig: Detectors_mountain} and \ref{Fig: Detectors_magnetic}, we observe that the GW frequency of isolated WDs could be 
much larger as compared to that of galactic binaries. This is because, in the case of isolated WDs, the spin frequency is responsible 
for the GW generation, whereas, in the case of binaries, their orbital periods are essential. Therefore, the confusion noise of the binaries 
does not affect the detection of the isolated WDs. However, the GW frequency of some other sources, such as massive binaries and compact 
binary inspirals, is similar to that of isolated WDs. Hence, using specific templates for binaries, the objects can be distinguished 
from each other. Of course, the frequency range of isolated WDs is such that neither the nano-hertz detectors, such as IPTA, EPTA, and 
NANOGrav, nor the other currently operating ground-based detectors such as LIGO, VIRGO and KAGRA, can detect them. 

From Figure \ref{Fig: Mass_radius plot}, it is evident that the GR dominated WDs are considerably 
big as compared to the $f(R)$ gravity dominated ones, particularly at higher 
central densities, provided they possess the same central density. 
Moreover, using the equations \eqref{Eq: h0 mountain} and \eqref{Eq: h0 magnetic}, we have $h_0 \propto I$ 
with $I$ being the typical moment of inertia of the body. As a result, if we compare two WDs with the 
same ellipticity and angular frequency, the GR dominated WD emits stronger gravitational radiation. Since 
this paper is dedicated to studying the effect of $f(R)$ gravity, we do not explicitly calculate $h_0$ 
in the case of GR, like we have calculated it in detail in our earlier papers \citep{2019MNRAS.490.2692K,2020IAUS..357...79K,2020ApJ...896...69K}. 
Moreover, we target to explore the detectability of those WDs which exhibit sub- and super-Chandrasekhar 
limiting mass WDs in the same $M-\mathcal{R}$ relation, which GR based theory cannot, and it has 
many consequences outlined in the Introduction. 
However, it is to be noted that $f(R)$ gravity dominated WDs, being smaller in size, can rotate much 
faster with frequency $\gtrsim 1$ Hz, which is not possible in the case of a regular WD governed by GR. 
In this paper, we have shown that in the presence of small deformation, such as the presence of a rough 
surface or magnetic fields, these WDs can emit intense gravitational radiation, which can later possibly 
be detected by BBO, DECIGO, ALIA, and Einstein Telescope.

\begin{figure}[!htpb]
	\centering
	\includegraphics[scale=0.5]{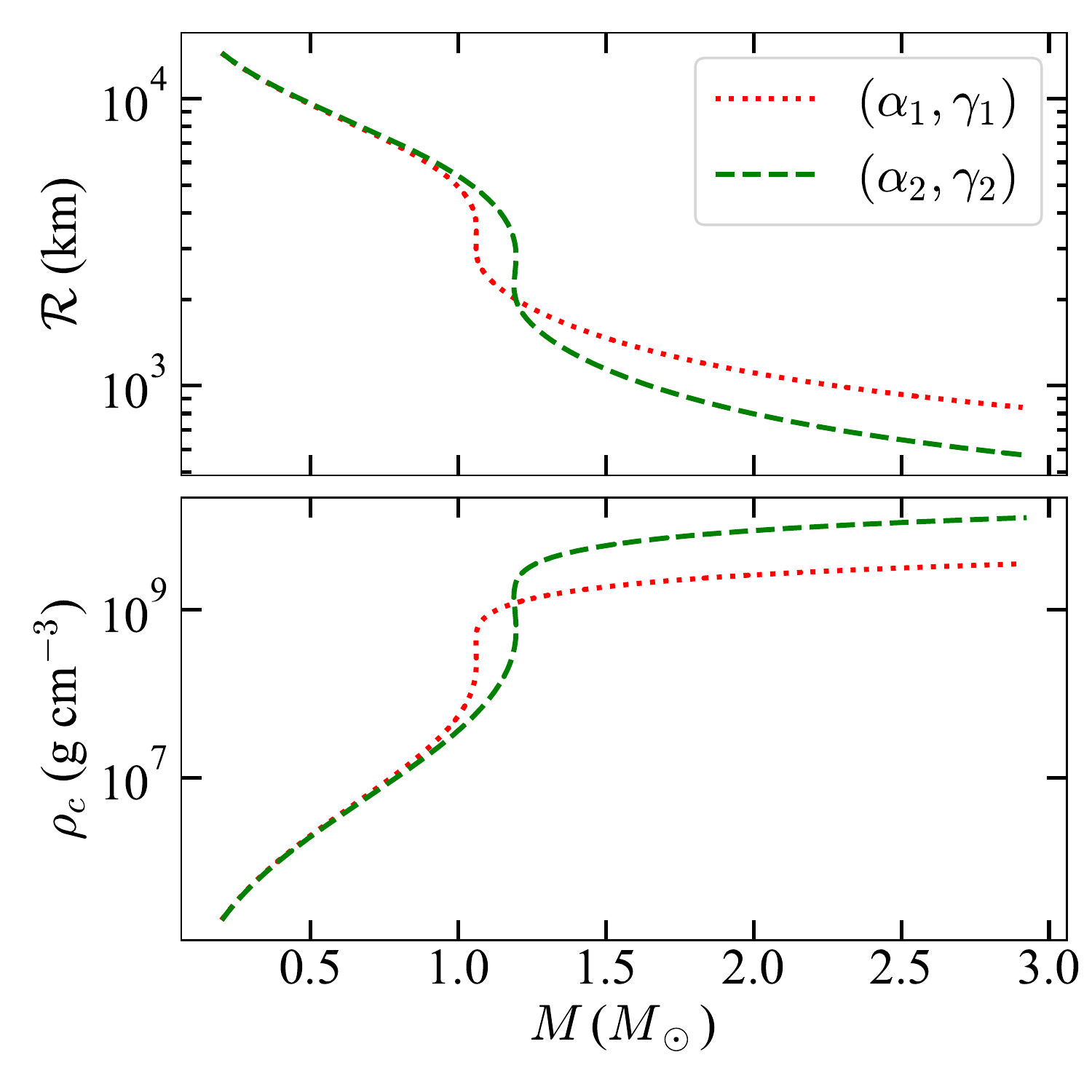}
	\caption{$M-\mathcal{R}$ and $M-\rho_c$ relations for WDs for 
	$f(R)=R+\alpha R^2(1-\gamma R)$ model. The values of $\alpha$ and $\gamma$ in the units of cm$^2$ are $(\alpha_1, \gamma_1) = (10^{14},10^{17})$ and $(\alpha_2, \gamma_2) = (3\times 10^{14},3\times 10^{17})$.}
	\label{Fig: New mass radius}
\end{figure}

The birth rate of He-dominated WDs is $\sim 1.5\times 10^{-12}$ pc$^{-3}$ yr$^{-1}$ \citep{1983Ap&SS..97..305G}, which means within 
100 pc radius, only one WD is formed in approximately $10^6$ yrs. Hence, continuous GW from some massive 
WDs, which have radiation timescales (life span) $\sim 10^{8-9}$ yrs, can be detected, as shown in the 
Figures \ref{Fig: Detectors_mountain} and \ref{Fig: Detectors_magnetic}. If the advanced futuristic 
detectors, such as DECIGO, BBO, or Einstein Telescope detect the isolated WDs, one can quickly check 
whether the physics is governed by GR, or $f(R)$ gravity, or any other modified theory of gravity as 
follows. Once the GW detectors detect such a WD, we have the information of $h_0$ and 
$\Omega_\text{rot}$. Consequently, if the distance to the source $r$ is known by some other method, then 
by using the Equations \eqref{Eq: h0 mountain} and \eqref{Eq: h0 magnetic}, one can estimate the 
ellipticity and, thereby, can predict the mass and size of the WD. This will be a direct detection of 
the WD with low thermal luminosity (usually a super-Chandrasekhar WD which is smaller in size). In this 
way, we obtain the exact mass--radius relation of the WD. Since different theories provide different 
mass--radius relations of the WD, by obtaining the exact mass--radius relation from observation, one can 
rule out various theories from the zoo of modified theories of gravity which do not follow this 
relation. For example, Figure \ref{Fig: New mass radius} shows mass--radius relations for a particular $f(R)$ gravity model but with different sets 
of model parameters. Comparing these results with those shown in Figure 
\ref{Fig: Mass_radius plot} reveals that the range of radius for stable 
WDs depends on the
chosen model parameters for the same ranges of mass and central density. Direct detection of the WDs will provide valuable information to identify the correct radius range and hence mass--radius relation of the WDs.


\section{Conclusions} \label{Conclusions}

In this paper, we have established a link between theory and possible GW observations of the WDs in a $f(R)$ gravity. Various 
researchers have already proposed hundreds of modified theories of gravity, including many $f(R)$ gravity theories, and each one of them 
possesses its own peculiarity. However, because of the lack of advanced observations near extremely high gravity regime, nobody, so far, 
can rule out most of the models to single out one specific theory of gravity. In this paper, we consider one valid class of $f(R)$ 
gravity model from the solar system constraints, which can explain both the sub- and super-Chandrasekhar WDs, along with the normal WDs, depending only on their central 
densities keeping the parameters of the model fixed. However, from the point of observation, the primary difficulty is that we do not 
know the size of the peculiar WDs and, hence, the exact mass--radius curve is still unknown. Thereafter, we calculate the strength of GW 
emitted from these WDs, assuming they slowly rotate with little deformation due to some other factors, such as the presence of roughness 
of the envelope, or the presence of a weak magnetic field. If the advanced futuristic GW detectors, such as ALIA, DECIGO, BBO, or 
Einstein Telescope can detect these WDs, one can estimate the ellipticity of the WDs and, thereby, put bounds on the WD's 
size. This will restrict the mass--radius relation of the WD, which can rule out various modified theories, and we will be inching 
towards the ultimate theory of gravity.

\acknowledgements
The authors would like to thank Clifford M. Will of the University of Florida for his insightful comments on the effect of modified gravity in tri-axial systems. S. K. would also like to thank Khun Sang Phukon of Nikhef, Amsterdam, for the useful discussions on integrated SNR in the case of various detectors. Finally, thanks are	due to the anonymous referee for a thorough reading the manuscript and comments that have helped to improve the presentation of the work, particularly comments on the mass--radius relations of various models. B.M. acknowledges a partial support by a project of Department of Science and Technology (DST), India, with Grant No. DSTO/PPH/BMP/1946 (EMR/2017/001226).


\bibliographystyle{yahapj}
\bibliography{bibliography}

\end{document}